\documentclass[journal]{IEEEtran}
\usepackage[utf8]{inputenc} 
\usepackage[T1]{fontenc}    
\usepackage{hyperref}       
\usepackage{float}
\usepackage{graphicx}
\usepackage[caption=false]{subfig}
\usepackage[font={small,it}]{caption}
\usepackage{multirow}

\usepackage{enumitem}
\usepackage{booktabs} 
\usepackage[normalem]{ulem}
\newcommand{\phs}{\textit{Surface}}

\newcommand{\phh}{\textit{Hand Held}}

\newcommand{\spyker}{{Spearphone}}
\newcommand{\gendCl}{\textit{Gen-Class}}
\newcommand{\spkCl}{\textit{Spk-Class}}
\newcommand{\sphRec}{\textit{Speech-Class}}
\newcommand{\voice}{\textit{Voice scenario}}
\newcommand{\mm}{\textit{Multimedia scenario}}
\newcommand{\ass}{\textit{Assistant scenario}}
\begin{document}

\title{Motion Sensor-based Privacy Attack on Smartphones}

\author{S Abhishek Anand, Chen Wang, Jian Liu, Nitesh Saxena, Yingying Chen 
	\thanks{S Abhishek Anand and Nitesh Saxena are with the University of Alabama at Birmingham. Email:$\left\{ {anandab, saxena} \right\}$@uab.edu. }
	\thanks{Chen Wang, Jian Liu and Yingying Chen are with WINLAB, Rutgers University. Email:$\left\{ {chenwang, jianliu} \right\}$@winlab.rutgers.edu, $yingche$@scarletmail.rutgers.edu. }
}
\pagestyle{plain}

\maketitle

\begin{abstract}
In  this paper, we build a speech privacy attack that exploits \textit{speech reverberations} generated from a \textit{smartphone's inbuilt
loudspeaker}\footnote{Inbuilt loudspeakers are different from the earpiece speaker that is used to listen to incoming calls} captured via a
zero-permission motion sensor (accelerometer).  We design our attack \textit{\spyker}\footnote{\textbf{Spearphone} denotes \textbf{S}peech
\textbf{p}rivacy \textbf{e}xploit via \textbf{a}ccelerometer-sensed
\textbf{r}everberations from smart\textbf{phone} loudspeakers}, and demonstrate
that speech reverberations from inbuilt loudspeakers, at an appropriate
loudness, can impact the accelerometer, leaking sensitive information about the
speech.  In particular, we show that by exploiting the affected accelerometer
readings and carefully selecting feature sets along with off-the-shelf machine
learning techniques, \spyker\ can successfully perform \textit{gender
classification} (accuracy over 90\%) and \textit{speaker identification}
(accuracy over 80\%) for any audio/video playback on the smartphone. Our results with testing the attack on a voice call and voice assistant response were also encouraging, showcasing the impact of the proposed attack. In addition, we perform \textit{speech recognition}
and \textit{speech reconstruction} to extract more information about the
eavesdropped speech to an extent.
Our work
	brings to light a fundamental design vulnerability in many
	currently-deployed smartphones, which may put people's speech privacy
	at risk while using the smartphone in the loudspeaker mode during
	\textit{phone calls}, \textit{media playback} or \textit{voice
	assistant interactions}.

\end{abstract}

%


\vspace{-2mm}

\vspace{-2mm}
\section{Introduction}
\label{sec:intro}
\vspace{-1mm}
Today's smartphones contain a plethora of sensors aiming to provide a
comprehensive and rich user experience. Some common sensors used in modern
smartphones include infrared, accelerometer and gyroscope, touchscreen, GPS,
camera and environmental sensors.  A known security vulnerability associated
with smartphone motion sensors is the unrestricted access to the motion sensor
readings on most current mobile platforms {(e.g., the Android OS)}, essentially
making them \textit{zero-permission} sensors.  Recent research
\cite{ALHAIQI2013989, anandspeechless, cai2011touchlogger,
Mantyjarvi2005Gait,marquardt2011sp, michalevsky2014gyrophone} exploits motion
sensors for eavesdropping on keystrokes, touch input and speech. {Since the
Android mobile operating system has a market share of 75.16\% worldwide and
42.75\% in the United States  \cite{gs}, this security vulnerability is of extreme concern
especially in terms of speech privacy.}

Expanding on this research line in significant ways, we investigate a new
attack vulnerability in motion sensors that arises from the \textit{co-located}
speech source on the smartphone (smartphone's in-built loudspeakers). Our work
exploits the motion sensors (accelerometer) of a smartphone to capture the
speech reverberations (surface-aided and aerial) generated from the
smartphone's loudspeaker while listening onto a voice call or any media in the
loudspeaker mode. These speech reverberations are generated due to the smartphone's body vibrating due to the principle of forced vibrations \cite{Coleman1988}, behaving in a manner similar to a sounding board of a piano. Using this attack, we show that it is
possible to compromise the speech privacy of a live human voice, without the
need of recording and replaying it at a later time instant.

As the threat of exploiting smartphone's loudspeaker privacy using motion
sensor arises due to co-location of the speech source, i.e., the phone's
loudspeaker, with the embedded motion sensors, it showcases the perils to a
user's privacy in seemingly inconspicuous threat instances, some examples of which are
described below:

\begin{itemize}
		[leftmargin=*]
	\item \textit{Remote Caller's Speech Privacy Leakage in Voice Calls}: The proposed attack can eavesdrop on voice calls to compromise the speech privacy of a remote end user in the call. A smartphone's loudspeaker can leak the speech characteristics of a remote end party in a voice call via its motion sensors. These speech characteristics may be their gender, identity or the spoken words during the call (by performing speech recognition or reconstruction).
    \item \textit{Speech Media Privacy Leakage:} In the proposed attack, on-board motion sensors can also be exploited to reveal any audio/video file played on the victim's smartphone loudspeaker. In this instance, the attacker could exploit motion sensors, by logging the output of motion sensors during the media play, and learn about the contents of the audio played by the victim. This fact could also be exploited by advertisement agencies to spam the victim by using the information gleaned from the eavesdropped media content (e.g., favorite artist).
    \item \textit{Voice Assistant Response Leakage}: Our proposed threat may extend to phone's smart voice assistant (for example, Google Assistant or Samsung Bixby), that communicate with the user by reaffirming any given voice command using the phone's loudspeakers. While this action provides a better user experience, it
also opens up the possibility of the attacker learning the voice assistant's responses.
\end{itemize}

Considering these attack instances, we explore the vulnerability of motion sensors to speech reverberations, from the smartphone's loudspeakers, conducted via the smartphone's body. We also examine the frequency response of the motion sensors 
and the hardware design of the smartphones that leads to the propagation of the 
speech reverberations from the phone's loudspeaker to the embedded motion sensors.
Our contributions are three-fold:


\begin{enumerate}[leftmargin=*]	
	\item \textbf{\textit{A New Speech Privacy Attack System:}} We propose a novel attack, \spyker\ (Section \ref{sec:threatModel}), that compromises speech privacy by exploiting the embedded motion sensor (accelerometer) of a smartphone. Our work targets speech reverberations (surface-aided and aerial vibrations), produced by the smartphone's loudspeakers, rather than the phone owner's voice, which is directed towards the phone's microphone. This includes privacy violation of remote caller on a voice call (live at remote end but still played through phone owner's loudspeakers), user behavior by leaking information about audio/video played on phone's loudspeakers or the smartphone's voice assistant's response to a user query (including the issued command) through the loudspeakers in a preset voice. Accelerometers are not designed to sense speech as they \textit{passively reject air-borne vibrations} \cite{Coleman1988}. 
	
	Thus, it is very hard for an attacker to eavesdrop on speech using accelerometer readings. Indeed, prior work on motion sensor exploits for compromising speech required the speech to be replayed via \textit{external loudspeakers} while a \textit{smartphone (with embedded motion sensors) was placed on the same surface as the loudspeaker}. In contrast, our work leverages the speaker inbuilt in the smartphone to provide a fundamentally different attack vector geared towards eavesdropping on \textit{speech reverberations}	(a detailed comparison with prior work is provided in Section \ref{sec:background}).
    \spyker\ is a three-pronged attack that performs gender, speaker and 
		speech classification using accelerometer's response to the speech reverberations, generated by the victim's phone's speakers. 

    \item \textbf{\textit{Attack Design and Implementation:}} As a pre-requisite to the \spyker\ attack, we perform frequency response analysis of motion sensors (accelerometer and gyroscope) to determine
    the sensor most susceptible to our attack (Section \ref{sec:sensorDesign}). We find accelerometer to be the most receptive and therefore design our attack based on its readings associated with smartphone loudspeaker's speech signals. The attack is designed to work on the Android platform, facilitated due to the ``zero-permission'' nature of motion sensors {(up to the latest Android 10)}.    
    We execute the attack by carefully using off-the-shelf machine learning and 
    signal processing techniques (Section \ref{sec:design}). By using known techniques and tools, we believe that our attack implementation has a significant value as it can be created by low-profile attackers. Although we use standard methods to keep our attacks more accessible, we had to address several technical challenges like low sampling rates of the motion sensors and appropriate feature set selection as discussed below and in Section \ref{sec:challenges}.
    
	\item \textbf{\textit{Attack Evaluation under Multiple Setups:}} We evaluate \spyker\  
    under multiple setups mimicking near real-world usage of smartphone loudspeakers (Section \ref{sec:attack}). 
    We show that Spearphone can perform gender and speaker classification on media playback, requiring as low as just one word of test data with an f-measure $\ge$0.90 and $\geq$0.80 respectively, which shows the threat potential of the attack. Promising, although slightly lower, classification results are obtained for the voice
call and voice assistant response scenarios. The speech classification result also shows the possibility of speech identification, essentially turning it into a loudspeaker for the attacker. Our evaluation and datasets capture the three threat instances as they all require the speech signals to be output by the phone's loudspeakers.
\end{enumerate}

\vspace{-3mm}
\section{Background and Prior Work}
\label{sec:background}
\vspace{-1mm}
The embedded motion sensors (i.e., accelerometer and gyroscope) are useful in supporting various mobile applications that require motion tracking or motion-based command. However, they also bring potential risks of leaking user's private information.
Due to the nature of the motion sensors, they can capture the vibrations associated with users' movements such as typing on the phone's keyboard. This could cause sensitive information leakage on mobile devices~\cite{miluzzo2012tapprints,cai2011touchlogger, owusu2012accessory,marquardt2011sp,xu2012taplogger}. 
For instance, TouchLogger~\cite{cai2011touchlogger}, TapLogger~\cite{xu2012taplogger} and Accessory~\cite{owusu2012accessory} utilize the accelerometer and gyroscope embedded on smartphones to infer keystroke sequence or passwords when the user inputs on the smartphone's keyboard. 
TapPrints~\cite{miluzzo2012tapprints} further shows that the tap prints on the smartphone touchscreen can be characterized by its accelerometer and gyroscope to identify users. 
In addition, (sp)iPhone~\cite{marquardt2011sp} shows the vibrations generated by typing on a physical keyboard can be captured by a nearby smartphone's accelerometer to derive the user's input. 

Additionally, it is necessary to consider speech privacy in various daily scenarios (e.g., private meetings, phone conversations, watching or listening to media). In order to prevent unintentional listeners from overhearing the 
speech, traditional methods apply sound-proof walls for a closed conference room to confine the speech within the room.
Besides, microphone access on a smartphone is subjected to a high-level permission to prevent adversarial exploits. To prevent potential snooping via smartphone's built-in microphone, people can simply deny any app's microphone permission if they are not actively using a specific feature that requires microphone.
The motion sensors, on the other hand, are usually freely accessible, meaning any application needs \textit{zero permission} to access them. Additionally, MEMS sensor attributes and structures could be affected by sounds, indicating the potential to leak the smartphone user's private speech information.

Existing studies have shown that background noise affects MEMS sensors and degrades their accuracy~\cite{castro2007influence,dean2007degradation,dean2011characterization}. 
The reason is that the MEMS structure can be resonant to some parts of frequencies of the sound vibrations surrounding the phone. 
However, due to the low sampling rate of motion sensors (e.g., $200$Hz on most smartphones), its capability of snooping speech sound is often ignored or underestimated. 
However, the recent work shows that embedded MEMS motion sensors could reveal speech information~\cite{michalevsky2014gyrophone,zhang2015accelword,anandspeechless}.
Specifically, Gyrophone~\cite{michalevsky2014gyrophone} shows that gyroscope is sensitive enough to measure acoustic signals from an external loudspeaker to reveal speaker information.
Accelword~\cite{zhang2015accelword} uses smartphone's accelerometer to extract signatures from the live human voice for \textit{hotwords} extraction. 

Speechless~\cite{anandspeechless} further tests the necessary conditions and setups for speech to affect motion sensors for the speech leakage. \cite{anandspeechless} concluded that motion sensors may indeed be influenced by external sound sources as long as the generated vibrations are able to propagate along the surface to the embedded motion sensors of the smartphone, placed on the same surface (\textit{surface-aided}). \cite{anandspeechless} also showed that \textbf{aerial vibrations of speech}, such as those produced by the vocal tract of live human speakers speaking in the microphone of the phone, \textbf{do not impact its motion sensors}. Pitchin \cite{pitchin} presented an eavesdropping attack using embedded motion sensors in an IoT infrastructure (having a higher sampling rate than a smartphone motion sensor) that is capable of speech reconstruction. They leveraged the idea of time-interleaved analog to digital conversion by using a network of motion sensors, effectively boosting the information captured by the motion sensors due to increased sampling rate obtained by sensor fusion.

The above studies, however, focus on studying the possibility or necessary
conditions for making the embedded motion sensors respond to the external sound
sources (e.g., loudspeaker and live human voice).  Our work explores the
possibility of revealing the speech played by the smartphone's built-in
speakers from the phone's own motion sensors. This setting is related to a
large number of practical instances, whose privacy issues are still unexplored.

Compared to the related work in \cite{michalevsky2014gyrophone}, we found that
the accelerometer performs much better than the gyroscope when picking up the
speech reverberations.  Moreover, the study in \cite{michalevsky2014gyrophone}
examined the speech from an external loudspeaker, which produces much stronger
sound/vibration signals and only targeted the local speaker's speech using
their smartphone. Smartphone loudspeakers lack the wide range of frequency
response compared to an external loudspeaker (with woofers), especially at low
frequencies. Since the speech signals that produce vibrations consist of low
frequencies, our threat model is much weaker than the one used in
\cite{michalevsky2014gyrophone}. Our work is not restricted to
just surface-aided speech vibrations as it exploits both surface-aided and aerial vibrations that are propagated within the
smartphone's own body.
Thus, we believe  \cite{michalevsky2014gyrophone} presents a threat model that
is extremely favorable to the attacker but potentially too restrictive to work
in the real world. A summary of related work vs. our work is provided in Appendix Table \ref{tab:comparison}.

In summary, in this paper, we identify and dissect live speech and media instances in which the speech privacy attack through motion sensors works, whereas a recent study \cite{anandspeechless} concluded these sensors to be ``speechless'' in most other setups (e.g., humans speaking into the phone, or when the loudspeaker does not share the same surface as the phone). {Our proposed setup and previous studies, such as \cite{anandspeechless} and \cite{michalevsky2014gyrophone}, use a similar scenario, but the key difference lies in the targeted speech. In previous works, the targeted speech was from sources external to the smartphone while we consider the speech that originates from the smartphone itself via speech reverberations. Our approach allows us to extend the threat impact to scenarios that were not considered in previous works. 
}  
We elaborate our detailed attack model in Section~\ref{sec:threatModel}.

\vspace{-2mm}
\section{Sensors vs. Speech Reverberations}
\label{sec:sensorDesign}
\begin{figure}
	\centering
	\includegraphics[scale=0.37]{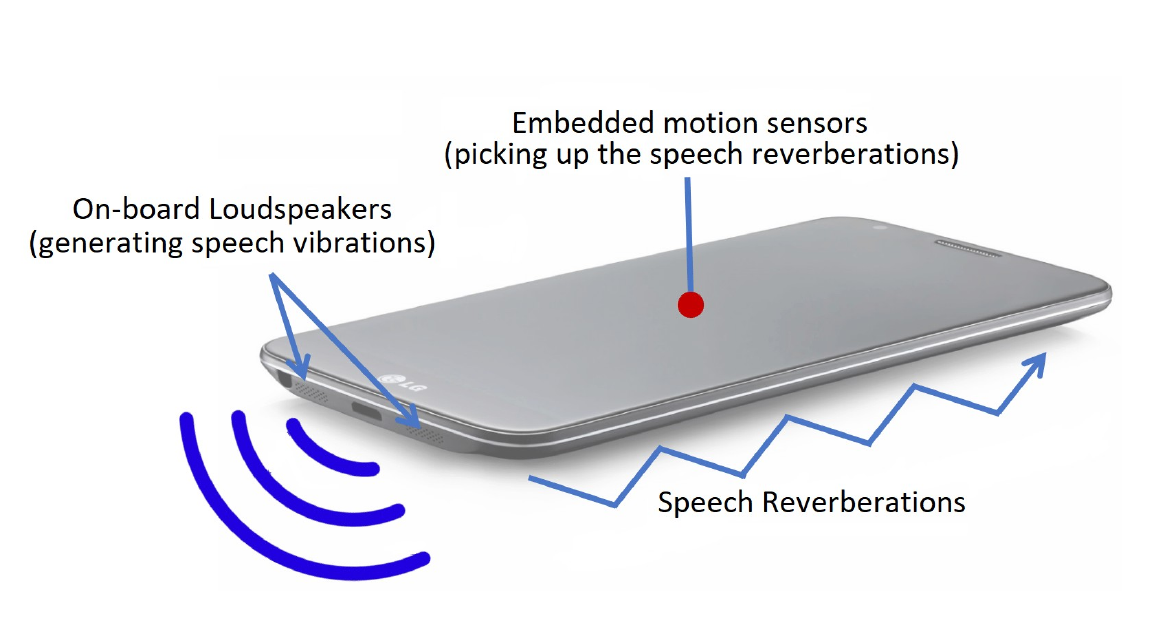}
	\vspace{-5mm}
	\caption{Speech reverberations, propagating within the smartphone's body, impact the motion sensors}
	\vspace{-6mm}
	\label{fig:illustrative}
\end{figure}

\vspace{-1mm}
As shown in Figure~\ref{fig:illustrative}, when a sound is played by a mobile phone, the phone 
loudspeakers generate sound vibrations.
Compared to the vibrations being transmitted in the air (\textit{air-borne propagation}),
the phone body also provides a pathway for propagating the resulting sound reverberations to the accelerometer and gyroscope, which are embedded in the phone (Figure \ref{fig:illustrative}). 
These embedded motion sensors are 
designed for sensing the phone motions, enabling various 
applications (e.g., fitness tracking) but they also suffer from exploitation (due to \textit{zero-permission} nature) that draws serious security and privacy concerns.




\vspace{-2mm}
\subsection{Accelerometer Frequency Response}
\label{sec:accelRes}
An accelerometer is an electro-mechanical
device for measuring acceleration  which can be either 
static (e.g., gravity) or dynamic (e.g., movement/vibration).
The MEMS accelerometer can be modeled as a mass-spring system. An
external acceleration force causes movement of the tiny seismic mass inside 
fixed electrodes, causing a capacitive electrical signal change which can be 
measured as acceleration value~\cite{kaajakari2009practical}.

Spearphone aims to capture the speech from the smartphone's built-in loudspeaker by leveraging the readily available motion sensors. To measure the frequency response of the accelerometer to the built-in loudspeaker sound, we play a specifically-designed signal and collect the accelerometer readings with a smartphone (e.g., Samsung Galaxy Note 4). The smartphone sensor sampling rate is set to its maximum limit of $250$Hz and it is placed on a wood table. We generate a chirp sound signal, sweeping from frequency $0$Hz to $22$kHz for $5$ minutes, and play the sound through the smartphone's built-in loudspeaker at maximum volume. 

The accelerometer amplitudes in Appendix Figure~\ref{fig:frequency_response1}(a) show that the accelerometer has a strong response to the sound frequency ranging from $100$Hz to $3300$Hz. This is because the built-in loudspeaker and the accelerometer are on the same device, and the sound gets transmitted through the smartphone components causing vibrations. Moreover, the spectrogram in Appendix Figure~\ref{fig:frequency_response1}(b) shows that different frequency sounds cause responses at the low frequency points of the accelerometer and generate aliased signals~\cite{michalevsky2014gyrophone}, which can be expressed by the equation $f_{a} = |f - N\cdot f_{s}|$, where $f_{a}$, $f$, $f_{s}$ are the vibration frequency of the accelerometer, sound frequency and the accelerometer sampling rate. $N$ can be any integer. Therefore, the accelerometer can capture rich information from the sound but with aliased signals in low frequency. 

To determine the dominant propagation medium in our proposed attack, we performed an experiment to compare the phone's accelerometer response in two settings: (1) an LG G3 phone's accelerometer captures the speech from its own loudspeaker; and (2) the G3 phone's accelerometer captures the speech from the loudspeaker of another phone placed near it on the shared table. The sound pressure level of the played speech is adjusted at the same level, and the distance between the loudspeaker and the G3 phone's accelerometer is kept the same. Appendix Figure~\ref{fig:propagation_medium} shows the root mean square (RMS) of the captured sensor readings. We observe that our attack setting (\textit{smartphone body}) possesses a much higher response, as high as $0.2$, than the \textit{shared solid surface} setting (around $0.05$). It indicates that the smartphone body dominates the vibration propagation so as to carry more speech-relevant information in the captured accelerometer readings.

\vspace{-2mm}
\section{Attack Overview and Threat Model}
\label{sec:threatModel}
In this section, we will describe \spyker\ threat model and provide an overview 
of \spyker~(also depicted in Appendix Figure \ref{fig:threat}) that showcases the motion sensor exploiting speech reverberations.
The threat model is based on \cite{michalevsky2014gyrophone,anandspeechless} where the embedded smartphone 
motion sensor readings are recorded in presence of speech in multiple setups.

\vspace{-2mm}
\subsection{\spyker\ Threat Instances}
\label{sec:threat_sc}
In \spyker, we assume that the smartphone's loudspeaker is being used to output any audio. Some examples of \spyker\ threat instances are described as follows:
\begin{itemize}[leftmargin=*]
    \item \textit{Voice Call: }In this threat instance (Figure \ref{fig:scene1}), the victim is communicating with another 
    person and listening in the \textit{loudspeaker mode} (i.e., not using the earpiece speaker or headphones). 
    We assume the phone loudspeaker is at the 
		maximum loudness level to produce strongest speech reverberations (although we also test the effect of lower volumes and validate the threat under such conditions). The phone could be hand-held or placed on a solid surface like a table. In this threat instance, the attacker is able to capture reverberations on the victim's phone, generated in real time during the phone call.
    
     \item \textit{Multimedia: }We also believe that the live call instance could extend to situations where human speech is produced by smartphone's loudspeakers while playing a media file.  While the content of the media may not be private, an attacker can get some confidential information about the victim (for example, Snapchat videos, preferred music). {Advertisement companies could use this information to target victims with tailor-made ads, inline with the victim's preferences.} Malicious websites can also track the motion sensor data output in background while media is played in the foreground. It could be a breach of privacy if a person's habits or behavior patterns are exposed to the attacker that could potentially be used against the victim to discriminate them from jobs, insurance purposes, financial benefits, etc. This threat instance is depicted in Figure \ref{fig:scene2}.
    
    \item \textit{Assistant: }Most modern smartphones come with an inbuilt voice assistant for performing intelligent tasks. The voice assistant 
    often confirms the user's command to ensure the desired action. It makes the process user-friendly and gives the user a choice to modify or cancel the current process. If the phone assistant uses the inbuilt phone loudspeakers, any response from the phone assistant is played back via these loudspeakers and can potentially affect the motion sensors, in turn exposing the intent of the user to an attacker exploiting the motion sensors (Figure \ref{fig:scene2}).
\end{itemize}

\vspace{-4mm}
\subsection{Attacker's Capabilities}
\label{sec:threat_cap}
The attacker in our threat model has similar capabilities as elaborated in 
previous literature \cite{anandspeechless,michalevsky2014gyrophone}. 
The attacker can fool the victim into installing a malicious application or a malicious website could track the motion sensor readings in the background via JavaScript while the unsuspecting victim is browsing. Michalevsky et al.\cite{michalevsky2014gyrophone} analyzed the sampling rate of motion sensors, as permissible on various browser platforms and found out that only Gecko-based browsers (e.g. Firefox) do not place any additional limitations on the sampling rate of the motion sensors. Thus, the malicious attack through Javascript on a Gecko-based browser would work similar to a malicious application installed on the Android platform.
These malicious applications can be designed to get triggered 
for specific threat instances described previously and can start logging the 
motion sensor output. The output can then be transmitted to the attacker where the attacker can extract confidential information.

The degree of threat posed by our attacker in \spyker~is measured by the extent of breach in speech privacy. \spyker~ attempts to compromise speech privacy by performing 
gender, speaker, and speech classification. From an attacker's perspective, gender classification helps the attacker to narrow down the set of speakers for unidentified speech samples thereby increasing the recognition accuracy for speaker identification. Speaker classification helps the attacker with more context about the communicated speech (in addition to revealing the identity of one of the parties involved in a private voice call) while speech classification reveals the contents of the speech itself that may be considered private between the two communicating parties. More specific privacy concerns for each type of classification/leakage are provided below. We also limit our threat model to utilize a finite set of words (a closed dictionary) although it could be expanded by identifying individual phonemes contained in the speech.

\begin{itemize}[leftmargin=*]
    \item \textit{Gender Classification (Gen-Class):} Gender classification can cause a privacy compromise in scenarios where the gender of a person may be used to target them in a harmful manner. For example, advertising sites could push spam advertisements of products aimed towards a specific gender \cite{datta2015AdPrivacy}. It can also be used to discriminate against a particular gender as shown in \cite{datta2015AdPrivacy} where job search advertisements were gender biased. Certain oppressive societies put restrictions on particular genders and may use gender classification to target individuals in potentially harmful ways. Gender classification is relevant to the \textit{Voice Call} threat instance (Section \ref{sec:threat_sc}) where the attacker is interested in the identity of the remote caller that can be narrowed down by the gender of the caller.
\spyker\ extracts speech features from unlabeled motion sensor recordings and classifies each extracted sample as originating from either male or female speaker using classification models built on previously obtained labeled samples of sensor recordings.
    
    \item \textit{Speaker Classification (Spk-Class):} Speaker classification involves identifying a speaker that could lead to privacy leakage of the communicating parties in a voice call. For example, an attacker can learn if a particular individual was in contact with the phone owner at a given time. Another example could be a person of interest under surveillance by law enforcement who is in contact with the phone owner. It could also lead to leakage of the entire phone log of the phone owner. This classification is most suited for the \textit{Voice Call} threat instance (Section \ref{sec:threat_sc}) where the identity of the remote caller can directly be revealed.
Similar to \gendCl\, the attacker builds a classification model from the labeled dataset of sensor recordings that associate each sample with a unique 
speaker and then tests the obtained unlabeled sensor recording against this model.
    
    \item \textit{Speech Classification (Speech-Class):} \spyker\ aims to learn
    the actual words transmitted via the phone's speaker during the attack. To perform \sphRec, 
    we build a classification model based on a finite word list. Speech features from the obtained sensor readings for isolated words 
    are compared against the labeled features of the word list by the classification model that 
    provides the attacker with a possible rendition of the actual spoken word. We also study the feasibility of performing speech reconstruction by isolating words from natural speech and then using word recognition on isolated words to reconstruct speech.
    
    Speech classification is relevant to all the three threat instances discussed in Section \ref{sec:threat_sc}. It can disclose the specifics of the call for the voice call threat scenario, leak the contents of the media consumed by the victim and reveal the actions taken by the voice assistant in response to the victim's commands.

\end{itemize}

\vspace{-5mm}
\subsection{Attack Setup}
\label{sec:threat_set}
The environment of the victim plays an important role in our threat model. In 
our model, we study the speech reverberations generated from the smartphone's inbuilt speakers.
Therefore, we exclude any external vibration generating source such as external loudspeakers
 studied in \cite{anandspeechless,michalevsky2014gyrophone}.
Our threat model assumes the victim's phone is the only device that is present in the 
environment and the only vibrations present in the environment are generated by 
the victim's smartphone speakers. This assumption is supported by the discoveries of \cite{anandspeechless} and \cite{Coleman1988} that indicated that aerial vibrations from ambient noise are too weak to affect the MEMS accelerometers.
To test the threat instances, we categorize two setups where the victim's 
phone speaker can impact the embedded motion sensors.
\begin{itemize}[leftmargin=*]
    \item \phs\ Setup: In this setup, the phone is kept on a flat surface with its screen
    facing up. This setup may be used in \voice\ where the victim places 
    the phone on a table while talking to someone with the phone on speaker mode.
    This setup also mimics occurrences when the phone is put on a table, countertop etc. in \mm\ and \ass.This setup is similar to \cite{michalevsky2014gyrophone} and \cite{anandspeechless} that were primarily focused on surface-borne propagation of speech vibrations via a shared surface. However, our setup also allows the possibility of aerial propagation of speech vibrations due to the very close proximity of the speech source (the phone speaker) and the accelerometer. As both reside within the same device body, we do not rule out the effect of aerial vibrations of speech on the accelerometer and hence use the encompassing term ``reverberations'' as indicated in Figure \ref{fig:illustrative}.
    
    \item \phh\ Setup: The victim may also hold the phone in hand while in \voice,
    playing a media file in \mm\ or using their phone's assistant in \ass. In our 
    threat model, we assume that while holding the phone in hand, the victim is stationary with no 
    hand or body movement. 
\end{itemize}

Lastly, the attacker in this threat model is not in the physical 
vicinity of the intended victim. The attack happens through a previously 
installed malicious application or a rogue website  that records the motion sensor data output 
during relevant time and sends it to the attacker. The attacker can examine the 
captured data in an off-line manner and use signal processing along with 
machine learning to extract relevant information about the intended victim.


%

\vspace{-2mm}
\section{Attack Design}
\label{sec:design}
\vspace{-1mm}

\spyker~uses a malicious application installed on the victim's phone (or through JavaScript running in a browser on the phone) to record motion sensor readings while the phone is on speaker mode. The malicious application is triggered only when the speakerphone is turned on on the victim's phone. {In the Android API's AudioManager class, we can use isSpeakerphoneOn() function to check if the speakerphone functionality is enabled.}

Spearphone relies on the loudspeakers of the smartphone to generate reverberations from received speech signals. We tested the ear piece speaker, that is normally used to listen to incoming phone calls (a target for our attacker). Appendix Figure \ref{fig:earLG} shows the spectrum of the accelerometer log,  recorded in the presence of an incoming voice call, that used the ear piece speaker. The call volume was set at maximum and the phone was placed on a solid surface. Appendix Figure \ref{fig:earLG} does not show any footprints of speech, indicating incapability of the ear piece speaker on LG G3 to produce speech reverberations strong enough to impact the accelerometer.

\vspace{-3mm}
\subsection{Motion Sensor Recording}
We designed an Android application that mimics the behavior of a malicious attacker (Section \ref{sec:threatModel}). On start, the application immediately begins logging motion sensor readings. After a delay of five seconds from the start, we play a single word on a separate thread in the application while it is recording motion sensor data. This step partially mimics the act of the callee's speech generated during a phone/voice call or the playing of a media file on the phone via the inbuilt loudspeakers. Our use of isolated words can also be extended to continuous speech, but we do not aim to implement a complete speech recognition system limiting only to showcase the threat posed by embedded motion sensors. Upon completion, we process the output file containing motion sensor readings as detailed in subsequent subsections.

\vspace{-2mm}
\subsection{Identifying Speech Areas}
Once the attacker obtains motion sensor output from the malicious application, he needs to extract speech areas for performing \gendCl, \spkCl\ and \sphRec\ as per Section \ref{sec:threat_cap}. Since we used isolated words in our attack, each speech sample contains one instance of a spoken word. As gyroscope did not display a noticeable presence of speech in the spectrum of its readings (Section \ref{sec:sensorDesign}), accelerometer is the only  motion sensor that is considered in \spyker. To extract speech from accelerometer recordings, we trim off the beginning five seconds and ending two seconds of the recordings to compensate for the initial delay before playing the isolated word and the ending finger touch for pressing the ``Stop'' button to pause the motion sensor recordings. 

Since we see maximum response along the Z axis, for accelerometer's reaction against speech (Section \ref{sec:sensorDesign}), we try to determine the speech areas along the Z axis readings and use corresponding areas for the X and Y axes. To determine the area of speech in the Z axis readings for accelerometer, a sliding window (size=100 samples) is used. Since different words have varying lengths of utterance, we picked the duration of the shortest word as the size of sliding window. We calculate variance in each window to determine the sensor behavior within that time. A higher variance in the readings indicates the presence of an external motion (speech vibrations). We extract the bounds of window with maximum variance as the sensor readings influenced due to the presence of speech.

\subsection{Feature Set for Speech Classification}
\label{subsec:featureset}
Once we have extracted accelerometer readings that contain speech, we need speech features for \gendCl, \spkCl\, and \sphRec\..
Mel-Frequency Cepstral Coefficients (MFCC) are widely used in audio processing as they give a close representation of human auditory system. While MFCC features are sensitive to noise, our threat model (Section \ref{sec:threatModel}) assumes minimal interfering noise. 

Time-frequency domain features are another option to classify a signal. These features consist of statistical features of the signal in time domain such as 
minimum, maximum, median, variance, standard deviation, range, absolute mean, CV (ratio of standard deviation and mean times 100), skewness, kurtosis, first, second and third quartiles, inter quartile range, mean crossing rate, absolute area, total absolute area, and total signal magnitude averaged over time. Frequency domain features are calculated by converting accelerometer readings from time domain to frequency domain using Fast Fourier transformation (FFT). The FFT coefficients were used to derive energy, entropy and dominant frequency ratio that are used as frequency domain features in time-frequency features.

\vspace{-2mm}
\subsection{Evaluation Metrics}
\label{sec:metrics}

We use the following metrics to evaluate the performance of \spyker~attack: Precision, Recall, and F-measure.
\textit{Precision} indicates the proportion of correctly identified samples to all the samples identified for that particular class. In other words, it is the ratio of number of true positives to number of elements labeled as belonging to the positive class. \textit{Recall} is the proportion of correctly identified samples to actual number of samples of the class. It is calculated as the ratio of number of true positives to number of elements belonging to the positive class. \textit{F-measure} is the harmonic mean of precision and recall. For perfect precision and recall, f-measure value is 1 and for worst, it is at 0.

\vspace{-2mm}
\subsection{Design Challenges}
\label{sec:challenges}

\subsubsection{Low Sampling Rates}
Operating Systems like Android place a hard limit on the data output rate for motion sensors, in order to conserve the battery life of the device. This behavior  helps in freeing in valuable processing and memory power. This fact, however, makes it harder to turn the on-board motion sensors into acting as a microphone for capturing speech. Compared to an audio microphone with a sampling rate ranging from 8kHz to 44.1kHz, motion sensors become severely limited in their sampling rate (120Hz on LG G3, 250Hz on Samsung Galaxy Note 4). In addition, the on-board loudspeakers may be limited in their capacity of reproducing the audio in its true form resulting in several missing frequencies outside the loudspeaker's range. Thus, we need to choose the motion sensor that can capture most of the speech signal. We compared the frequency response of both accelerometer and gyroscope in Section \ref{sec:sensorDesign}. The accelerometer response in Section \ref{sec:accelRes} shows us that it was able to register motion (acoustic vibrations) for the audio frequency range $100-3300$Hz. Comparing with gyroscope's response in Appendix Section \ref{sec:gyroRes}, we see that the gyroscope's response is considerably weaker than accelerometer in the human speech frequency range. Thus, we make use of accelerometer in our experiments.  

\subsubsection{Feature Set Selection}

We compared both MFCC and time-domain frequency features to determine the most suitable feature set accurately classifying the speech signals, captured by the accelerometer. We use the metrics (Section \ref{sec:metrics}) and the following classifiers: Support Vector Machine (used in \cite{michalevsky2014gyrophone}) with Sequential Minimal Optimization (SMO), Simple Logistic, Random Forest and Random Tree (variants of decision tree classifier used in \cite{zhang2015accelword}).
An initial experiment was conducted using the TIDigit word list \cite{tidigitsSub} for using isolated words on LG G3 smartphone in \phs~scenario. Our results indicated that time-frequency features outperformed MFCC features using 10-fold cross validation for all four classification algorithms. This result, combined with the fact that time-frequency features were proven to be efficient in \cite{zhang2015accelword}, led us to decide upon using it in our attack for \gendCl, \spkCl, and \sphRec. Among the classifiers, we noticed random forest outperforming other classifiers using the time-frequency features, hence we use it in the rest of our experiments (Appendix Figure \ref{fig:PhsTD} and Figure \ref{fig:PhsAm}). A full set of our time-frequency features is provided in Appendix Table \ref{table:feature_list}. 

\noindent {\textbf{Salient Time-frequency Features:}
We further studied the distribution differences of time-frequency features for \gendCl, \spkCl, and \sphRec, as not all the features exhibit the same capability for differentiating the sound during classification. Appendix Figure~\ref{fig:spearphone_salient_features} shows the distribution of a subset of the most salient features in box plots, which works best for \gendCl. In particular, the identified feature set includes the second quartiles (Q2), third quartiles (Q3), signal dispersion (SigDisp), mean cross rate (MCR), ratio of standard deviation over mean (StdMeanR) and energy, along different axes. Similarly, we also identified the most effective time-frequency features for \spkCl, and \sphRec~({boxplots presented in Appendix Figures \ref{fig:salient_feature_speaker} and \ref{fig:salient_feature_word}}).

\subsubsection{Complete Speech Reconstruction}
Performing speech reconstruction with the information captured by a low sampling rate and low fidelity motion sensors may not be sufficient to recognize isolated words. Moreover, it is unrealistic to generate a complete dictionary (i.e., training profile) of all the possible words for the purpose of user's full speech reconstruction. To address these issues, we extracted the time-frequency features from the accelerometer readings, which exhibit rich information to distinguish a large number of words based on existing classifiers (e.g., Random Forest and Simple Logistic). 
We performed word isolation by analyzing the spectrogram obtained from accelerometer readings under natural speech and calculated the Root Mean Square of the power spectrum values. We developed a mechanism based on searching the keywords (e.g., credit card number, targeted person's name and SSN) and only used a small-sized training set to reveal more sensitive information while ignoring the propositions, link verbs and other less important words.

\vspace{-2mm}
\section{Attack Evaluation}
\label{sec:attack}
\vspace{-1mm}
\subsection{Experiment Setup}
\label{sec:expsetup}
\noindent \textbf{Smartphones:}
We conducted our experiments using three different
smartphone models: LG G3, Samsung Galaxy S6 and Samsung Note 4.
The experiments were performed in a quiet graduate student laboratory on a table with hardwood top for \phs\ setup,  
while the \phh\ setup was conducted by two participants holding
the phone in their hands, which contains hand and body movements.
\begin{itemize}[leftmargin=*]
\item \noindent {\textit{Operating System:}}
We focused on phones with the Android mobile operating system as it does not require explicit user permission to obtain access to motion sensor data. In contrast, the iOS mobile operating system (from version 10.0 onwards) requires any application wishing to access motion sensor data to state its intent in the key ``NSMotionUsageDescription". The text in this key would be displayed to the user describing why the application wants to access the motion sensor data. Failure to state its intent in the above described manner results in immediate closure of the said application. Also, as pointed out in Section \ref{sec:intro}, the sizeable market share of Android (worldwide and the US) allows us to treat the threat posed to smartphones operating on this platform with extreme concern.
\item \noindent \textit{Sensors: } The accelerometer embedded in the smartphones used in our experiments had an output data rate of 4-4000 Hz and an acceleration range of $\pm2/\pm4/\pm8/\pm16$g. The liner acceleration sensitivity range are 0.06/0.12/0.24/0.48 mg/LSB. A quick comparison with the LSM6DSL motion sensor chip used in the latest Samsung Galaxy S10 smartphone indicates similar properties for the accelerometer.
 \vspace{-1mm}
\end{itemize}
\noindent \textbf{Word Datasets: }
\noindent \textit{TIDigits Dataset: } We used the subset of TIDigits corpus (\cite{tidigitsSub}). It contains $10$ single
digit pronunciation from ``0'' to ``9'' and $1$ additional
pronunciation ``oh''. It contains 5 male and 5 female speakers,
pronouncing the words twice. The sampling rate for the audio samples is 8kHz.

\noindent \textit{PGP words Dataset: } We also used a pre-compiled word list uttered by Amazon Mechanical Turk workers in a natural environment. The list
consisted of fifty-eight words from PGP words list and they were
instructed to record the words in a quiet environment. This data collection activity was
approved by the university's IRB and the participants had the choice to
withdraw from the experiment at any given time.
We used 4 male and 4
female Amazon Turk workers' audio samples (44.1 kHz sampling frequency). PGP word list is used for clear communication over a voice channel and is predominantly used in secure VoIP applications. 


\noindent \textbf{Speech Processing: }
We used Matlab for processing the accelerometer output
performing feature extraction as detailed in Section \ref{sec:design}. We used
Weka \cite{weka} as our machine learning tool to perform gender, speaker and speech classification on the extracted speech features. In particular, we test the attack with Random Forest classifier that outformed other classifiers as noted in Section \ref{sec:challenges}. We used default parameters for the classification algorithm and the detailed configurations are listed in Appendix Table~\ref{table:classifier_configuration}. We used both 10-fold cross-validation and the training and testing methods for classification. 10-fold cross-validation partitions the sample space randomly in 10 disjoint subspaces of equal size, using 9 subspaces as training data and retaining 1 subspace as testing data. For training and testing method, we split the dataset into training set and test set with the split being 66\% of the dataset being used for training and remaining 34\% being used for testing.

In our attack, the attacker collects the training samples for
building the classifier, which is unique for each device. Since our dataset is
not large (limited to 58 words for PGP words and 22 words for TIDigits), we
believe that it does not indicate a significant overhead for the attacker to procure the
training samples for each device targeted under the attack. Most other motion
sensor attacks to our knowledge (e.g., \cite{miluzzo2012tapprints,cai2011touchlogger,
owusu2012accessory,marquardt2011sp,xu2012taplogger, marquardt2011sp}), 
including Gyrophone, have similar or even more strict training requirements for
the attacker. 

\noindent \textbf{Effect of Noise: } In our threat model, the loudspeaker resides on the same device as the motion sensors thus any reverberations caused by the device's loudspeaker would impact the motion sensors. \cite{anandspeechless} and \cite{Coleman1988} claimed that external noise in human speech frequency range, traveling over the air, does not impact the accelerometer. Hence, any such noise in the surrounding environment of the smartphone would be unable to affect the accelerometer's readings.
The speech dataset used in our experiment, \textit{PGP words dataset}, was collected from Amazon Mechanical Turk workers, recording their speech in environments with varying degree of background noise. This dataset thus imitates the speech samples that the attacker may face in the real-world, such as during our attack instances involving phone calls.


    \vspace{-3mm}
\subsection{Gender and Speaker Classification in Voice Call Instance (\phs\ Setup) }
\begin{table}[]
\caption{Gender and speaker classification (10 speakers) for \phs\ setup using TIDigits and PGP words dataset using Random Forest classifier and time-frequency features}
\label{tab:Phs}
\scriptsize
\resizebox{\columnwidth}{!}{%
\begin{tabular}{c|c|c|c|c|}
\cline{2-5}
                                                 & \multicolumn{2}{c|}{\textbf{10-fold cross validation}} & \multicolumn{2}{c|}{\textbf{Test and train}} \\ \cline{2-5} 
                                                 & \textbf{TIDigits}         & \textbf{PGP words}         & \textbf{TIDigits}    & \textbf{PGP words}    \\ \hline
\multicolumn{5}{|c|}{\textbf{Gender classification}}                                                                                                     \\ \hline
\multicolumn{1}{|c|}{\textbf{Samsung Galaxy S6}} & 0.91                     & 0.80                      & 0.87                & 0.82                 \\ \hline
\multicolumn{1}{|c|}{\textbf{Samsung Note 4}}    & 0.99                     & 0.91                      & 1.00                 & 0.95                 \\ \hline
\multicolumn{1}{|c|}{\textbf{LG G3}}             & 0.89                     & 0.95                      & 0.85                & 0.95                 \\ \hline
\multicolumn{5}{|c|}{\textbf{Speaker classification}}                                                                                                    \\ \hline
\multicolumn{1}{|c|}{\textbf{Samsung Galaxy S6}} & 0.69                     & 0.70                      & 0.56                & 0.71                 \\ \hline
\multicolumn{1}{|c|}{\textbf{Samsung Note 4}}    & 0.94                     & 0.80                      & 0.92                & 0.80                 \\ \hline
\multicolumn{1}{|c|}{\textbf{LG G3}}             & 0.91                     & 0.92                      & 0.89                & 0.95                 \\ \hline
\end{tabular}}
\vspace{-5mm}
\end{table}

\subsubsection{Surface Setup using TIDigits}
The results for the \phs\ setup, where the victim's phone is placed on a surface such as a table, using TIDigits dataset is shown in Table
\ref{tab:Phs} for \gendCl\ and \spkCl.
We observe that the attack was able to perform \gendCl\ with a substantial degree of
accuracy f-measure $> 0.80$ with the attack being particularly successful on Note 4 as
demonstrated in Table \ref{tab:Phs}. 
As a baseline, the scores
are significantly better than a random guess attacker (0.50) indicating the
success of the attack in this setup.
For \spkCl, we note that the attack is more successful on LG G3 and Note 4 when
compared to Galaxy S6 with f-measure $>0.60$. 
A random guess attack performance is significantly worse at 0.10 (for 10 speakers) when
compared to this attack.
\subsubsection{Surface Setup using PGP words dataset} 

PGP words dataset results for \phs\ setup, are depicted in Table \ref{tab:Phs} for \gendCl\ and \spkCl. 
Comparing the attack against a random guess attack (0.50), we observe that the reported f-measure for the attack on all three phone models was more than 0.70 in both 10-fold cross-validation and train-test model. The attack on LG G3 boasted an f-measure of over $0.90$ consistently across all the tested classification algorithms leading to the conclusion that threat measure of \spyker\ when performing \gendCl\ may indeed be harmful in this setup.
Table \ref{tab:Phs} show \spyker's performance when \spkCl\ was performed using the PGP words dataset. 

For a 10-speaker classification model, a random guessing attack would give us an accuracy of 0.10. However, in our tested setup, we were able to achieve much higher f-measure scores with the attack on LG G3 achieving a score of almost 0.90. The attack on Galaxy S6 performed the worst among the attacks on all phone models but still had a better f-measure score of over 0.50 when compared to the baseline random guess attack accuracy. These results lead to conclusion that \spyker\ threat is also significant while performing \spkCl\ in this setup.

We also performed the binary classification for speakers by using two classes ``Targeted Speaker'' and ``Other'', that categorizes each data sample as either in the voice of the target speaker or any other speakers. We used PGP words dataset in our evaluation as it contained more words per speaker compared to TIDigits dataset. Using Random Forest classifier and 10-fold cross-validation, the mean f-score for this binary speaker classification for LG G3 was 0.97, for Galaxy S6 was 0.90, and for Note 4 was 0.94.

    \vspace{-2mm}
\subsection{Gender and Speaker Classification in Voice Call Instance (\phh\ Setup)}

\subsubsection{Hand-held Setup using TIDigits dataset} 

\begin{table}[]
\caption{Gender and speaker classification (10 speakers) for \phh\ setup using TIDigits and PGP words dataset using Random Forest classifier and time-frequency features}
\scriptsize
\label{tab:Phh}
\resizebox{\columnwidth}{!}{%
\begin{tabular}{c|c|c|c|c|}
\cline{2-5}
                                                 & \multicolumn{2}{c|}{\textbf{10-fold cross validation}} & \multicolumn{2}{c|}{\textbf{Test and train}} \\ \cline{2-5} 
                                                 & \textbf{TIDigits}         & \textbf{PGP words}         & \textbf{TIDigits}    & \textbf{PGP words}    \\ \hline
\multicolumn{5}{|c|}{\textbf{Gender classification}}                                                                                                     \\ \hline
\multicolumn{1}{|c|}{\textbf{Samsung Galaxy S6}} & 0.77                     & 0.72                      & 0.76                & 0.70                 \\ \hline
\multicolumn{1}{|c|}{\textbf{Samsung Note 4}}    & 0.81                     & 0.87                      & 0.77                & 0.88                 \\ \hline
\multicolumn{1}{|c|}{\textbf{LG G3}}             & 0.99                     & 0.95                      & 1.00                 & 0.95                 \\ \hline
\multicolumn{5}{|c|}{\textbf{Speaker classification}}                                                                                                    \\ \hline
\multicolumn{1}{|c|}{\textbf{Samsung Galaxy S6}} & 0.33                     & 0.34                      & 0.26                & 0.29                  \\ \hline
\multicolumn{1}{|c|}{\textbf{Samsung Note 4}}    & 0.73                     & 0.75                      & 0.61                & 0.70                 \\ \hline
\multicolumn{1}{|c|}{\textbf{LG G3}}             & 0.98                     & 0.93                      & 1.00                 & 0.95                 \\ \hline
\end{tabular}}
\vspace{-6mm}
\end{table}

Using the TIDigits dataset in \phh\ setup, we demonstrate the performance of the \spyker\ attack in Table \ref{tab:Phh} for \gendCl\ and \spkCl. For \gendCl, we observe that the performance of the attack on LG G3 is much better when compared to other devices for both 10-fold cross-validation model and train-test model with overall f-measure being approximately 0.70, which is again significantly better than a random guess attacker (0.50). 
For \spkCl, we see that the scores of Galaxy S6 are worse when compared to LG G3 with Note 4 having scores in between these devices. The f-measure values for LG G3 for \spkCl\ are over 0.90 for all the tested classifiers, for Note 4 these values are over 0.50 while Galaxy S6 values hover around 0.25. When compared to a random guess attack (0.10), the attack on G3 is significantly better while on Galaxy S6 it is slightly better. 

\subsubsection{Hand-held Setup using PGP words dataset} 
The \gendCl\ attack result is shown in Table \ref{tab:Phh}. The 10-fold cross-validation model indicates that the f-measure value of the attacker's classifier for LG G3 is the best performer among all three phone models. 
Similar to \phs, the attack performed better than a random guessing attacker (0.50) while the performance of attack was similar to the performance in \phs\ setup.
The attack's evaluation for \spkCl\ (Table \ref{tab:Phh}) shows that the attack is able to perform speaker identification with a high degree of precision for LG G3. The f-measure values, however, drop for Note 4 while the performance is worst for Galaxy S6. 
Thus, the attack's performance, while still better than a random guess attack (0.10), suffers a bit of setback for Note 4 and more so for Galaxy S6.
The binary classification for speakers (previously described in \phs\ setup) shows that the f-measure values when the smartphone is hand-held (\phh\ setup) are similar to \phs\ setup. The f-measure score averaged for 8 speakers with LG G3 was 0.97, for Galaxy S6 was 0.84, and for note 4 was 0.92.

\vspace{-2mm}
\subsection{Result Summary and Insights}
{The speaker classification accuracies for Note 4 and Galaxy S6 are higher for PGP words dataset compared to TIDigits dataset. This may be because PGP words dataset (sampled at 44.1kHz) was recorded at a higher sampling rate when compared to TIDigits (8kHz). This effect is not prominent in LG G3 because the sampling rate of its motion sensors is slightly lower (120Hz) than Note 4 or Galaxy S6 (around 200Hz).}
 {The gender and speaker classification accuracies seem to decrease a bit for the PGP words dataset in some instances. We believe that due to some background noise present in PGP words dataset, the accuracies may have been affected negatively. The accuracies of LG G3 do not seem to be impacted though, which we believe maybe due to its lower sampling rate (making it less prone to data degradation). }

Another interesting observation is that the \phs~setup overall produces better classification results than the \phh~setup. The hand motions and body movements are negative influences, but they only cause low frequency vibrations, which have been removed by our high-pass filter. Another possible explanation could be the vibration absorption/dampening caused by the holding hand. To test this reasoning, we conducted experiments with the Note 4 phone placed on a soft surface (i.e., soft couch). The gender classification accuracy is 87.5\%, similar to the handheld scenario (87\%), both of which are lower than the hard tabletop scenario. This suggests vibrations are possibly being absorbed by the hand to some degree. 
The speaker classification results overall seem similar to speaker classification using audio recordings \cite{khoury2013}. This behavior may be an indication that prominent speech features present in audio vibrations are also picked up by the accelerometer, as showcased by our experiments. 

Comparing our results with Michalevsky et al. \cite{michalevsky2014gyrophone}, we find that they achieved the best case gender classification accuracy of 84\% using DTW classifier on Nexus 4, which is lower than our best accuracy of almost 100\% using Random Forest classifier on Samsung Note 4, using the same dataset (TIDigits). For speaker classification, we obtained a higher accuracy of over 90\% using Random Forest classifier on Samsung Note 4 while that for Micalevsky et al. \cite{michalevsky2014gyrophone} was only 50\% for mixed gender speakers using DTW classifier for the same dataset (TIDigits). There is still room for improving the accuracy by exploring more features and deep learning methods, which will be explored in our future work.

\begin{table}[t]
	
	\caption{Effect of loudness on gender and speaker classification accuracy using Samsung Note 4 for \phs\ setup using Random Forest classifier and time-frequency features.}
	\label{tab:loudness}
	\scriptsize
	\resizebox{\columnwidth}{!}{%
		\begin{tabular}{cc|c|c|c|}
			\cline{3-5}
			&                                                                       & \multicolumn{3}{c|}{\textbf{Volume Level}}                                                                                \\ \cline{3-5} 
			&                                                                       & \textit{\textbf{$75\% Vol_{max}$}} & \textit{\textbf{$80\% Vol_{max}$}} & \textit{\textbf{$Vol_{max}$}} \\ \hline
			\multicolumn{1}{|c|}{\multirow{2}{*}{\textbf{\begin{tabular}[c]{@{}c@{}}Gender \\ Classification\end{tabular}}}} & \textit{\begin{tabular}[c]{@{}c@{}}TIDigits \end{tabular}}  & 0.93                                   & 0.90                                   & 0.99                                    \\ \cline{2-5} 
			\multicolumn{1}{|c|}{}                                                                                           & \textit{\begin{tabular}[c]{@{}c@{}}PGP word \end{tabular}}  & 0.78                                   & 0.95                                   & 0.91                                    \\ \hline
			\multicolumn{1}{|c|}{\multirow{2}{*}{\textbf{\begin{tabular}[c]{@{}c@{}}Speaker\\ Classification\end{tabular}}}} & \textit{\begin{tabular}[c]{@{}c@{}}TIDigits \end{tabular}}  & 0.45                                   & 0.70                                   & 0.94                                    \\ \cline{2-5} 
			\multicolumn{1}{|c|}{}                                                                                           & \textit{\begin{tabular}[c]{@{}c@{}}PGP words \end{tabular}} & 0.54                                   & 0.79                                   & 0.80                                    \\ \hline
	\end{tabular}}
	\vspace{-5mm}
\end{table}

\noindent {\textbf{Effect of Loudness:}}
We also evaluate the impact of the smartphone speaker volume on the performance of Spearphone. In particular, we test the gender and speaker classification performance of Spearphone when setting the smartphone speaker volume to 100\%, 80\%, and 75\% of the maximum volume. Table~\ref{tab:loudness} presents the results of the test on Samsung Note 4 phone, when it is placed on the table (i.e., \phs\ setup). The results show that while lower volume does impact the accuracy negatively, the lower volumes still achieve very high accuracy (i.e., 80\% volume achieves 95\% accuracy for gender classification and 79\% accuracy for speaker classification with the PGP words dataset). Also, the results indicate that the lower volume still causes significant privacy leakage, when compared to the random guessing accuracy (i.e., 50\% for gender classification and 10\% for speaker classification). 

People tend to use maximum volume to make the speech clear and comprehensible to avoid missing any important information~\cite{volume2019}. The louder volume, while providing clearer speech, would expose speech privacy more significantly via our Spearphone attack. In addition, we believe that the quality of the speakerphones on smartphones will improve over time and there are also powerful speaker cases in use today that can be physically attached to the phones \cite{big2019,polarpro2019}, and speech leakage over such higher quality speakerphones could be more devastating, even at lower volume levels.

\noindent {\textbf{Natural Speech Dataset:} While Spearphone achieves very high accuracy for the isolated word data set (i.e., TIDigits/PGP words), we further evaluated the performance of Spearphone with a more challenging natural speech dataset (VoxForge~\cite{voxforge}), which provides samples of sentences (10 words long on average) spoken by 5 male and 5 female speakers, with 100 samples for each speaker. In particular, for speaker classification, Spearphone achieves 91.3\% with LG G3 using Random Forest for 10-speaker classification under 10-fold cross validation. The result is very similar to the speaker classification with the isolated word datasets, which indicates that the attack is significant in a practical natural speech scenario. }

\noindent {\textbf{Realistic Voice Call Scenario:}
{To evaluate the threat of \spyker\ in more realistic scenarios like a real voice call, we downgraded the sampling rate of our PGP words dataset to 8kHz. The gender and speaker classification results (using f-measure scores) on Samsung Note 4 for the dataset using random forest classifier and 10-fold cross validation method were 0.73 and 0.47. For LG G3, the gender and speaker classification results measured as f-measure were 0.99 and 0.60. Compared to Table \ref{tab:Phs}, we see an expected drop in the speaker classification accuracy for the downsampled PGP words dataset. The gender classification accuracy degrades for Note 4 but such opposite behavior is observed for LG G3. }

\vspace{-4mm}
\subsection{Speech Recognition in Voice Calls}
\label{subsec:speech_recognition}
\vspace{-1mm}
We next demonstrate the feasibility of speech recognition using \spyker. We found that the G3 phone on a wooden table surface exhibited better performance when revealing speaker information. Towards this end, we utilized G3 on a wooden table to investigate the feasibility of \sphRec. We compared the performance of using time-frequency features with that of MFCC features, which are known to be popular in the speech recognition and found that time-frequency features give better classification accuracy than MFCC features. We also noted that random forest classifier outperformed the other tested classifiers, so we used Random Forest as our classifier on time-frequency features.
\begin{table}[t]
\centering
\caption{Speech recognition results for PGP words and TIDigits datasets using Random Forest classifier and time-frequency features on LG G3}
\label{tab:speech_rec}
\scriptsize
\resizebox{\columnwidth}{!}{%
\begin{tabular}{c|c|c|c|c|}
\cline{2-5}
\textbf{}                               & \multicolumn{2}{l|}{\textbf{10-fold cross validation}} & \multicolumn{2}{l|}{\textbf{Test and train}} \\ \cline{2-5} 
\textbf{}                               & \textbf{TIDigits}         & \textbf{PGP words}         & \textbf{TIDigits}    & \textbf{PGP words}    \\ \hline
\multicolumn{1}{|l|}{Single Speaker}    & 0.74                     & 0.81                      & 0.62                & 0.74                 \\ \hline
\multicolumn{1}{|l|}{Multiple speakers} & 0.80                     & 0.75                      & 0.71                & 0.67                 \\ \hline
\end{tabular}}
\vspace{-5mm}
\end{table}

\vspace{-1mm}

\subsubsection{\sphRec\ for Single Speaker}

\noindent \textbf{TIDigits dataset:}
Table \ref{tab:speech_rec} shows \spyker's accuracy of successfully recognizing a single speaker's $11$ isolated digit numbers (TIDigits dataset). For 10-fold cross validation, using time-frequency features, we achieved an f-measure score of $0.74$ with Random Forest classifier. In comparison, a random guess attacker would achieve an accuracy $0.09$ for the tested dataset. Similar results were obtained using train-test method for classification as in Table \ref{tab:speech_rec}, though there was a slight decrease in recognition accuracy.



\noindent\textbf{PGP words dataset:}
We further experimented with PGP words to explore how accurate \spyker\ could recognize the isolated words other than the digits. Table \ref{tab:speech_rec} shows the \sphRec\ results under 10-fold cross validation. By using the time-frequency features, \spyker\ achieved a much higher f-measure score of $0.81$ in recognizing words in a $58$-word list than digits. In comparison, the random guess accuracy was only $0.02$  for the dataset. The results of the train-test model showed a slight decrease in performance.

\subsubsection{\sphRec\ for Multiple Speakers}
There are plenty of scenarios involving multiple people's voices presenting on a single phone such as conference calls via Skype. 
We studied the feasibility of speech recognition from multiple speakers. In particular, we involve two speakers (one male; one female). Table \ref{tab:speech_rec} also shows the f-measure scores when recognizing digit numbers from the two speakers (multiple speaker scenario). We got an f-measure score of $0.80$ for the TIDigits dataset while the f-measure score for PGP words dataset, for multiple speaker scenario, was $0.75$. {We also used the PGP dataset, downsampled to 8kHz, to mimic real world telephony voice quality. The speech recognition accuracy for multiple speakers was $0.61$, which as expected, is lower than the original dataset but still above the random guess accuracy (i.e., 0.017).}


Gyrophone~\cite{michalevsky2014gyrophone} also carried out the speech recognition task by using TIDigits dataset and $44$ recorded words. However, they addressed a totally different attack setup where the sound sources were from an external loudspeaker and can achieve an accuracy of up to $0.65$. 
Our results of speech recognition accuracy around $0.82$ strongly indicate the vulnerability of smartphone's motion sensors to its own loudspeaker's speech. By combining the speech recognition and speaker identification, \spyker\ is capable of further associating each recognized word to the speaker identity in multi-speaker scenarios.

%
%

\vspace{-2mm}
\subsection{Speech Recognition in Multimedia and Voice Assistant Instances}
\label{subsubsec:speechMulti}
We also evaluated the \spyker\ accuracy in multimedia and voice assistant threat instances that are detailed in Section \ref{sec:threat_sc}. We used the same techniques that we used for \sphRec\ in voice calls (section \ref{subsec:speech_recognition}.
We simulated the multimedia threat instance by utilizing VoxCeleb dataset. VoxCeleb dataset \cite{voxceleb} is a large-scale audio-visual dataset of human speech, extracted from interview videos of celebrities uploaded to YouTube. We used a single speaker, 100 word dataset (where word truncation was done manually to extract each word) and the average length of a word in the 100 word dataset was 7.2 characters. Using random forest classifier on time-frequency features and 10-fold cross validation method, we were able to achieve a speech recognition accuracy of 0.35 for LG G3. The classification accuracy when the dataset was reduced to 58 words (for comparison with the PGP word dataset performance) was 0.54 for LG G3. Compared to the speech classification accuracy for PGP words dataset in Table \ref{tab:speech_rec} for LG G3, we see a decrease in the accuracy from 0.81 to 0.54. A random guess attack has an accuracy of 0.01 (for 100 words) and 0.017 (for 58 words) indicating that our attack outperforms it by an order of 30. We attribute the decrease in the classification accuracy to the existing noise in the Youtube recordings.

\begin{figure}[t]
	\centering
	\includegraphics[scale=0.33]{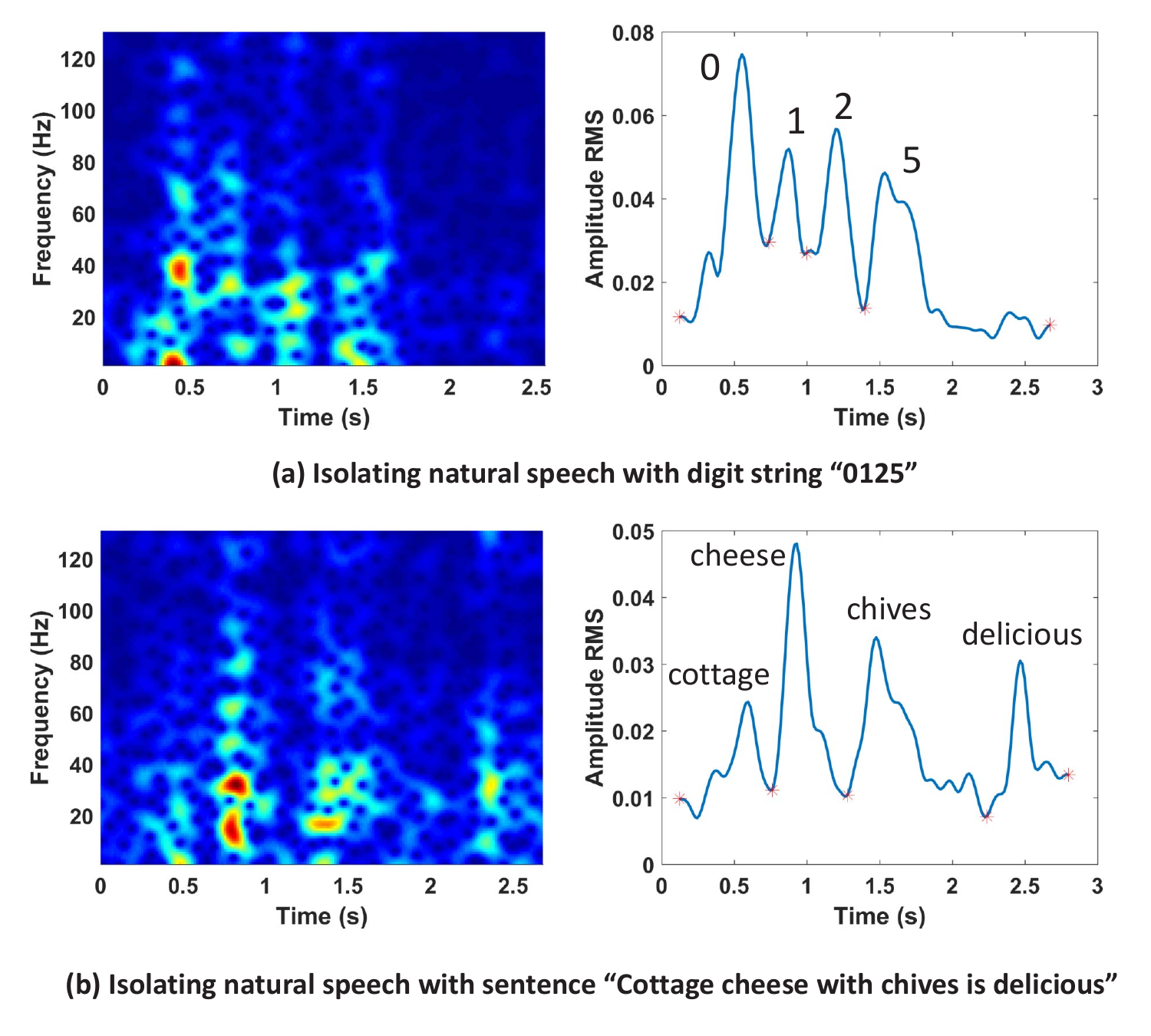}
	\vspace{-2mm}
	\caption{Illustration of the word isolation based on the RMS of the accelerometer spectrum}
	\label{fig:word_isolation}
	\vspace{-5mm}
\end{figure}
{We used Alexa voice assistant and generated PGP words dataset in Alexa's voice using the text-to-voice tool \cite{textvoiceio}. The text-to-voice tool pairs with the Alexa voice assistant and provides a text input feature to the user that is redirected to Alexa for repeating the user input. The speech recognition accuracy for the 58 words PGP word dataset was 0.31 for LG G3. This classification accuracy is again lower than the one reported in Table \ref{tab:speech_rec} for LG G3 on PGP words dataset, which was 0.81. Compared to a random guess attack, the proposed attack outperforms it by a magnitude of 18. However, Alexa's voice assistant is not human voice, albeit an artificially generated voice. Our feature set described in Section \ref{subsec:featureset} was tuned for recognizing characteristics of reverberations resulting from human voices. We propose reevaluating the feature set in our future work that is tuned based on artificially generated voices.}
\vspace{-2mm}
\subsection{Speech Reconstruction (Natural Speech)}
We have shown the capability of \spyker\ to recognize isolated words with high accuracy. 
To reconstruct natural speech, \spyker\ performs \textit{Word Isolation} and \textit{Key Word Search}, which first isolates each single word from the sequence of motion sensor readings and then searches for sensitive numbers/words from isolated words based on speech recognition introduced in Section~\ref{subsec:speech_recognition}. 

\subsubsection{Word Isolation}
\label{sec:speechIso}
In order to reconstruct natural speech, the words of the speech need to be first isolated from the motion sensor readings and then recognized individually.
However, isolating the words from the low sampling rate and low fidelity motion sensor readings is hard. To address this challenge, we calculated the Root Mean Square (RMS) of the motion sensor's spectrum at each time point and then located local peaks based on a pre-defined threshold to isolate each word. Figure~\ref{fig:word_isolation} illustrates an example of isolating a TIDigit string (``0125'') and a PGP sentence (``Cottage cheese with chives is delicious''). 
The motion sensor's spectrograms were converted to the amplitude RMSs at the right side of the figure. Based on the derived amplitude RMS, the valleys between the local peaks were detected to segment the critical words. 
We observed that some propositions and link verbs (e.g., ``with'' and ``is'') could hardly be detected, but this drawback has minimal effect on our results as these words do not affect the ability to understand an entire sentence.
We further evaluated our word isolation method by testing $20$ sentences containing around 28 words per sentence, and achieved $82\%$ isolation success rate. By excluding the less important propositions and link verbs, we achieve around $96\%$ success rate.   

\subsubsection{Key Word Search}
\begin{figure}[t]
\centering
\includegraphics[scale=0.35]{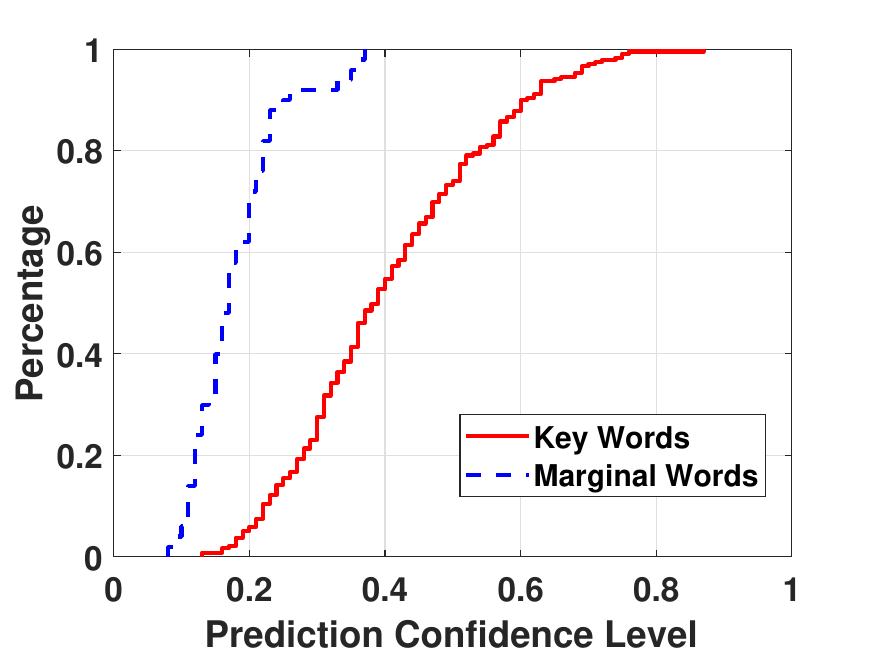}
	\vspace{-2mm}
\caption{CDF of the prediction confidence level for key words and marginal words}
\label{fig:keyword_search}
	\vspace{-7mm}
\end{figure}

Key word search is also significant when addressing natural speech. As it is hard to train all the potential words of a natural speech beforehand, the adversary might be more interested in the sensitive numbers/words (\textit{key words}) (e.g., credit card information, an important person's name, SSN, etc.) while marginal words such as propositions, link verbs and other such words can be ignored. Thus, a limited-size dataset is sufficient for stealing most sensitive information. 

After obtaining the isolated words, an adversary could search for key words based on a pre-constructed training model. In particular, \spyker\ relies on the predication probability returned by the training model as the confidence level to filter the key word search results. Figure~\ref{fig:keyword_search} shows the CDF of prediction confidence levels when $2/3$ PGP words are used as key words. We observed that the key words have higher confidence levels compared to marginal words. Thus, we could apply a threshold-based method to only focus on recognizing the keywords. Further combination of word isolation and key word search to reconstruct natural speech, requires fine-grained segmentation of the words and usage of Hidden Markov/other linguistic models for word corrections. This work is an avenue for possible future work.

\vspace{-1mm}
\section{Discussion and Future Work}
\label{sec:discussion}

\noindent \textbf{Attack Limitations: }In our experiments, we initially put the smartphone loudspeakers at maximum volume. Thus, the speech from the smartphone's loudspeakers was able to produce the strongest reverberations in the body of the smartphone, making maximum impact on the accelerometer. In reality, the loudness of different phones varies among different phone models and the loudness is also selective to each user. {Hence, we tested the effect of loudness on the attack's accuracy and found out that decreasing the volume from maximum to 80\% still allowed the attack to perform gender and speaker classification with significant accuracy, although at a lower accuracy compared to the full-volume attack.}

While our experiments tested two different datasets, they are still limited to single word pronunciations and are limited in size. However, single word accuracy can be extended to full sentence reconstruction using language modeling techniques. Moreover, TIDigits dataset, while relatively small, can still be effective in identifying sensitive information that mainly consists of digits. Personal information such as social security number, birthday, age, credit card details, banking account details etc. consist mostly of numerical digits. So, we believe that the limitation of our dataset size should not downplay the perceived threat level of our attack.


{Our attack targeted accelerometers embedded in the smartphones that were sensitive to the inbuilt speakers. The reverberations from the speakers travel via the body of the smartphone to the affected accelerometer. In most of the smartphones (including the smartphone models tested in this work), the motion sensor chip resides on the motherboard while the loudspeaker component is a separate unit \cite{ifixitNote4}. However, all of these components are fitted in the same device tightly to reduce the overall size (thickness) of the device, leading to reverberations traveling from the loudspeaker component to the motion sensor chip.}


Low sampling rate was a challenge during the implementation of our attack. Low sampling rate results in fewer data points collected by the motion sensors that directly impacts the accuracy of our attack. Resampling the obtained data to a higher sampling rate does not increase the amount of information contained in the collected data. To mitigate this challenge, we compared the accuracy of a combination of several feature sets and machine learning algorithms that maximized the amount of extracted information from our collected dataset.

Noise in the audio may be another limitation that would negatively impact \spyker\ classification results. We have tried to take the noise factor into account by using Amazon Turk workers to record our speech dataset that introduces a natural level of background noise in the recorded speech samples (in individual Amazon Turk worker's environment). Another factor to consider is the hand movement of the victim while holding the smartphone. Our attack experiment involved placing the phone either on a surface or held stationary in hand. Both these setups keep the smartphone stationary. However, they may not always be the case since the victim can move around with the smartphone or perform hand motions while holding the smartphone. Accelword \cite{zhang2015accelword} analyzed the impact of hand/body movements on accelerometers embedded in the smartphones and concluded that a cutoff frequency of 2 Hz would filter out the effect of these motions. Application of such a filter could make the proposed attack compatible with mobile setups, where the smartphone is not stationary.

\noindent 
\textbf{Impact of Hardware Design:}
\spyker\ uses the smartphone's accelerometer to capture the speech of the
inbuilt loudspeaker. However, the specific hardware designs of the smartphones of
various vendors are different, which results in the different capabilities of
the smartphone to capture the speech with accelerations. In particular, the
speaker properties and the accelerometer specifications are different across
various smartphone models. The specifications of the speaker and accelerometer
of the three popular smartphone models are summarized in Appendix
Table~\ref{table:smartphone_specification}. 

The accelerometers of the three models are similar but the loudspeaker of Galaxy S6 is less
powerful which may account for lower accuracy results on S6, especially in \phh\  where there is no contact between the smartphone's body and a solid surface so the reverberation effect may be reduced. Besides, the positions of the
speaker and the accelerometer on the smartphone may cause the acceleration patterns to respond to the same speech word differently. This is because the reverberations caused by the sound may transmit through different routes and get
affected by different complex hardware components. Appendix Figure~\ref{fig:hardware_reason} shows the motion sensor specifications for some popular brands of smartphones~\footnote{https://www.gsmarena.com/}. For example, the speakers of LG G3 and Note 4 are at the back of the smartphone, which can generate different levels of reverberations when placed on the table. In comparison, Galaxy S6's speaker is located at the bottom edge of its body, thereby having a diminished effect when placed on the table.

In this work, we focused on speech reverberations from the smartphone's loudspeakers as the source of privacy leakage. While previous works exploited speech vibrations from external speech sources, \spyker\ leverages the leakage of speech reverberations, that is possible due to forced vibration effect within the smartphone's body. These reverberations may be surface-aided or aerial, or a combination of both. A laser vibrometer could classify these reverberations, which will be our future work.

	\noindent \textbf{Accelerometer Models}:
	The three phone models tested in this paper are embedded with the Invensense accelerometer as summarized in Appendix
	Table~\ref{table:smartphone_specification}. We further analyzed the frequency response of another smartphone (Samsung Galaxy S3), embedding the STMicroelectronics accelerometer chip, to speech signals played via onboard loudspeaker. Our analysis suggests that the response is similar to the LG G3 (Invensense accelerometer) and both accelerometers show the frequency range between 300Hz and 2900Hz. This indicates that the STMicroelectronics accelerometer is picking up speech reverberations similar to the tested Invensense accelerometer. With the MEMS technology getting better and the loudspeakers being louder and more refined with every new generation of smartphone, we believe our attack should raise more concerns about speech privacy from this perspective.

\noindent \textbf{Potential Countermeasures:}
The design of any side channel attack exploiting motion sensors is
centered around the \textit{zero permission} nature of these sensors. To
mitigate such attacks, Android platform could implement stricter
access control policies that restrict the usage of these
sensors. In addition, users should be made aware of the implications of
permissions that they grant to applications. However, a stricter access control policy for the sensors directly affects the usability of the smartphones. Even implementing the explicit usage permission model by the applications often does not work since users do not pay proper attention to the asked permissions \cite{Felt2012}. They often do not read all required permissions, and even when reading, they are unable to understand the security implications of granting permissions. Moreover, many apps are designed to be overprivileged by developers \cite{Felt2011}. 

In addition, due to signal aliasing, vibrations of a wide range of frequencies are mapped non-linearly to the low sampling rate accelerometer data. Both the higher frequencies and lower frequencies contain the speech information. Thus, simply applying filters to remove the upper or lower frequencies cannot mitigate this attack.

A potential defense against \spyker\ could
also be set up by altering the hardware design of the phone. The internal build
of the smartphone should be such that the motion sensors are insulated from the
vibrations generated by the phone's speakers. One way to implement this
approach would be to mask or dampen the vibrations leaked from the phone's
speakers by surrounding the inbuilt speakers with vibration dampening material.
This form of speech masking would prevent speech reverberations emanated from the phone's speakers, possibly without affecting
the quality of sound generated by the speaker. Speaker isolation pads are already in use in music industry in recording studios for limiting sound vibration leakage \cite{speakerpad}. Other solutions like \cite{vibdamp} also exist that seek to dampen the surface-aided vibration propagation that may be useful in preventing leakage of speech vibrations within the smartphone. Further work is necessary to evaluate such a defensive measure against the threat studied in the paper.

\vspace{-2mm}
\section{Conclusion}
\label{sec:conclusion}
\vspace{-1mm}
We proposed a novel side-channel attack that compromises the phone's loudspeaker privacy by
exploiting accelerometer's output impacted by the emitted speech. This attack can leak information about the remote human speaker (in a voice call)
and the speech that is produced by the phone's speaker. In the
proposed attack, we use off-the-shelf machine learning and signal processing
techniques to analyze the impact of speech on accelerometer data and
perform gender, speaker and speech classification with a high accuracy. 

Our attack exposes a vulnerable threat scenario for accelerometer that
originates from a seemingly inconspicuous source (phone's inbuilt speakers). This threat can encompass several usage instances from daily activities like regular audio call, phone-based conference bridge inside private rooms, hands-free call
mode and voicemail/messages played on the phone. 
This attack can also be used to determine a
victim's personal details by exploiting the voice assistant's responses.
We also discussed some possible mitigation techniques that may help prevent such
attacks.
\vspace{-3mm}
\bibliographystyle{IEEEtran}
\bibliography{./bib/ms}

\clearpage
\appendix
\section{Appendix}
\subsection{Comparison with Prior Speech Privacy Motion-sensor Attacks}
 \begin{table}[H]
 	\vspace{-2mm}
 	\centering
 	\includegraphics[scale=0.6]{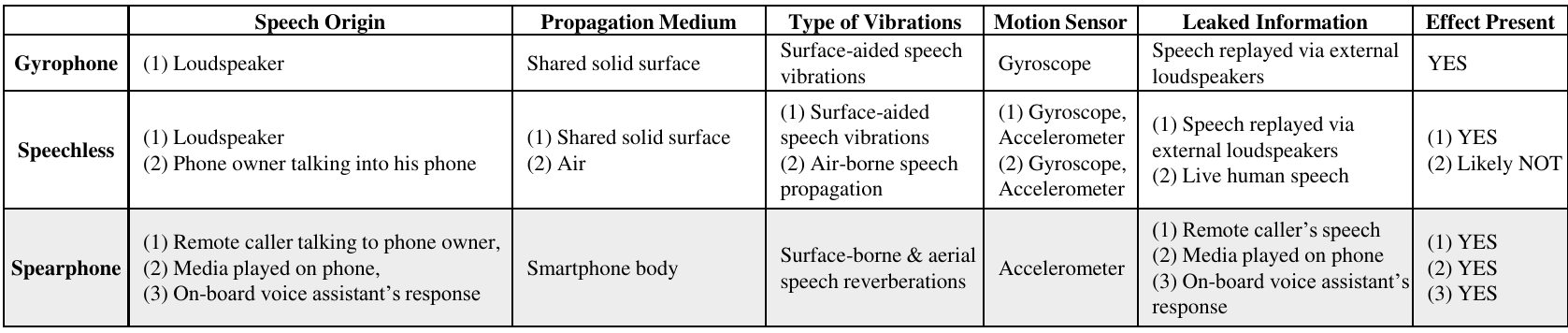}
 	\caption{\spyker~vs. prior speech privacy motion-sensor attacks/studies}
 	\label{tab:comparison}
 	\vspace{-2mm}
  \end{table}
  
  \subsection{Threat Model Overview}
 \begin{figure}[H]
	\centering
	\subfloat[Threat instance involving a voice call]{\includegraphics[width=3.2in]{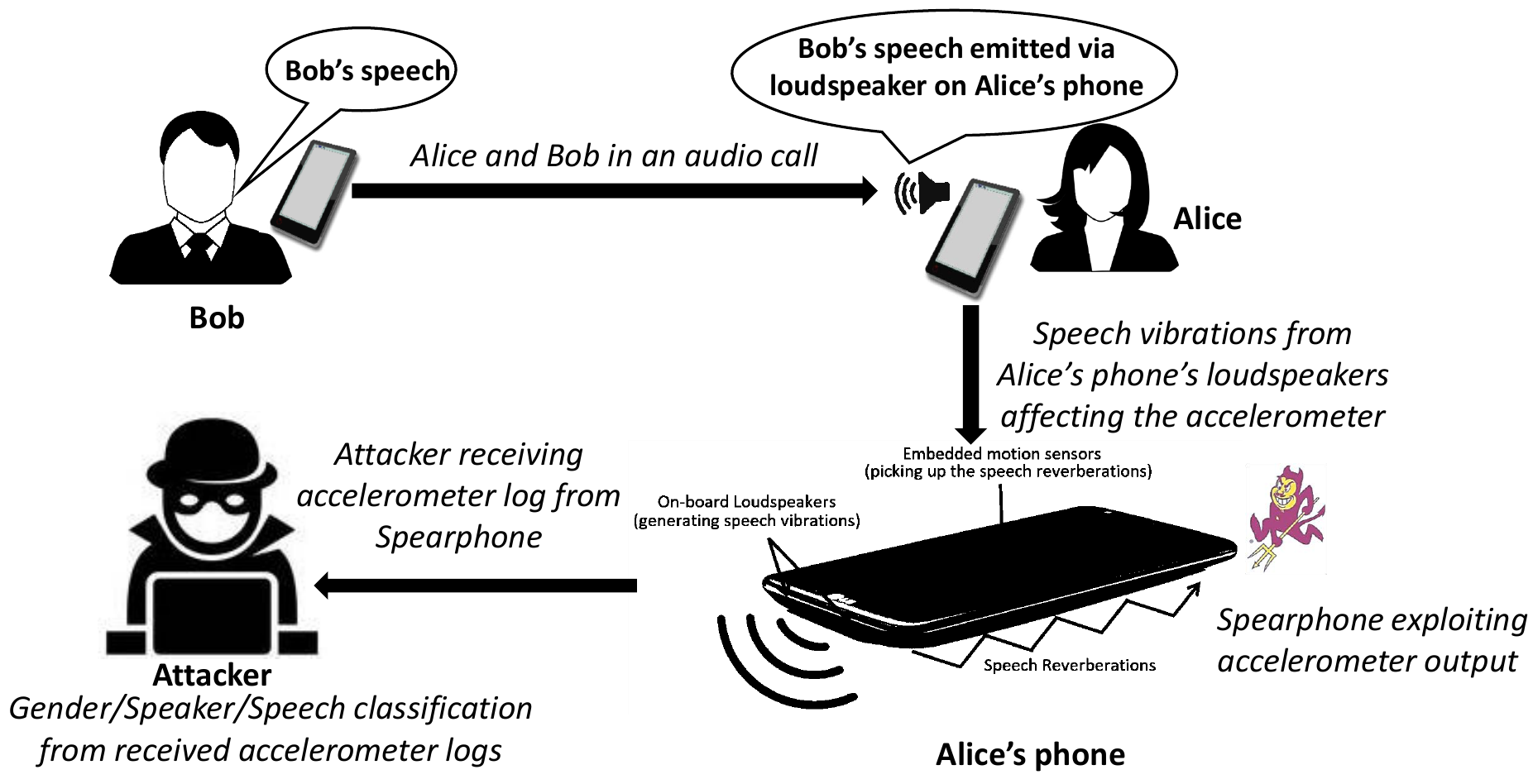}
		\label{fig:scene1}}
	\subfloat[Threat instance involving multimedia/voice assistant use]{\includegraphics[width=3.2in]{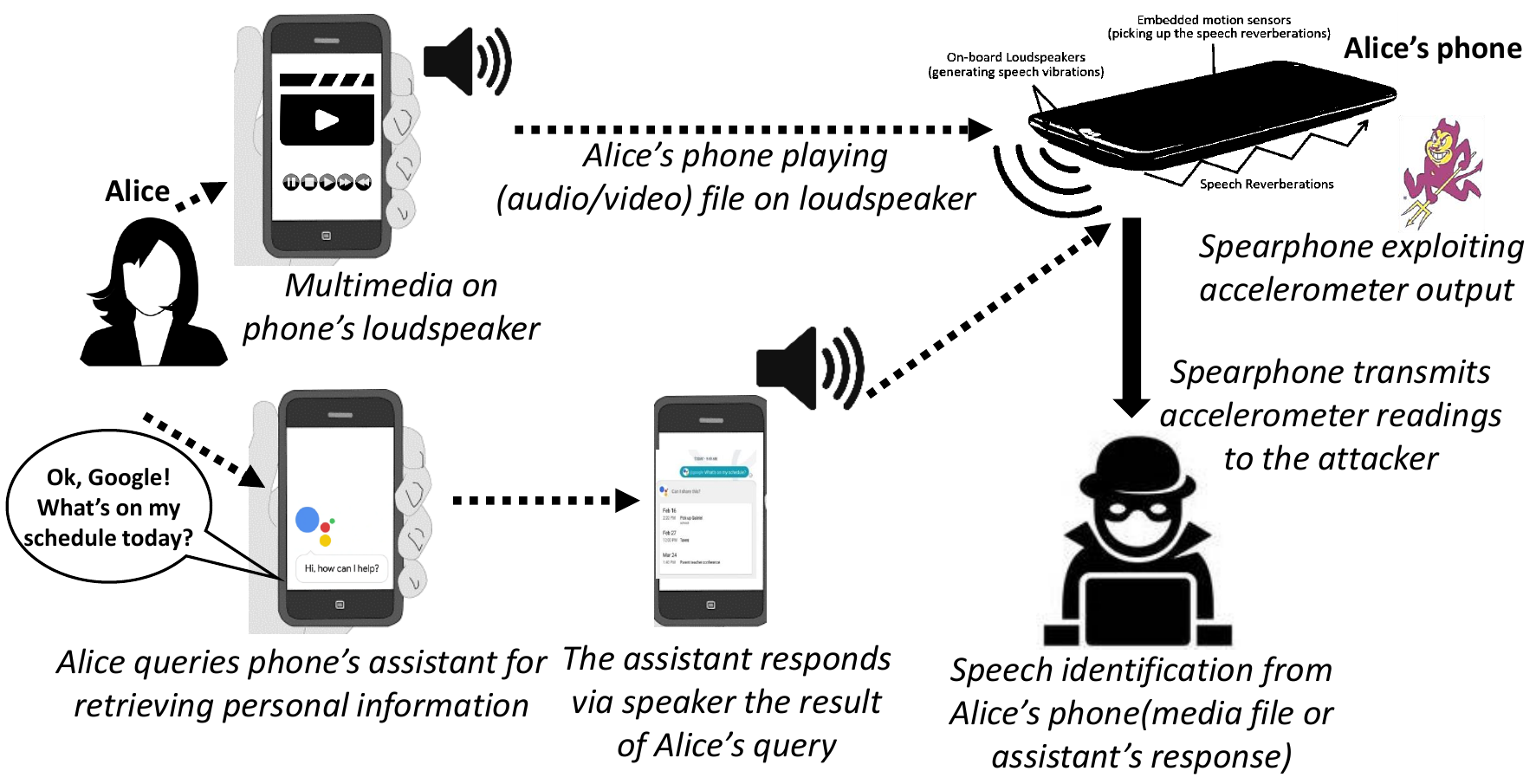}
		\label{fig:scene2}}
	\caption{An overview of the proposed attack depicting possible threat instances and the attack mechanism}
	\label{fig:threat}
\end{figure}

 \subsection{Classifier Configurations}
 \begin{table}[H]
 	\centering
 	\caption{Configurations of tested classifiers}
 	\label{table:classifier_configuration}
 	 \resizebox{\columnwidth}{!}{
 	\begin{tabular}{|l|l|}
 		\hline
 		Classifier & Configurations \\ \hline
 		SimpleLogistic & -I 0 -M 500 -H 50 -W 0.0 \\ \hline
 		SMO & \begin{tabular}[c]{@{}l@{}}-C 1.0 -L 0.001 -P 1.0E-12 -N 0 -V -1 -W 1 -K\\ -kernal PolyKernel -E 1.0 -C 250007\\ -calibrator Logistic -R 1.0E-8 -M -1 -num-decimal-places 4\end{tabular} \\ \hline
 		RandomForest & -P 100 -I 100 -num-slots 1 -K 0 -M 1.0 -V 0.001 -S 1 \\ \hline
 		RandomTree & -K 0 -M 1.0 -V 0.001 -S 1 \\ \hline
 	\end{tabular}}
 \end{table}
 
 \clearpage
 \subsection{Device Specifications}
 \begin{table}[H]
    \centering
    \caption{The specifications of the speakers and motion sensors for some popular brands of smartphones}
    \label{table:smartphone_specification}
    \resizebox{\columnwidth}{!}{
    \begin{tabular}{|c|c|c|c|}
        \hline
        \textbf{Smartphone}                                              & \textbf{\begin{tabular}[c]{@{}c@{}}Motion \\ Sensor\end{tabular}} & \textbf{Output Data Rate} & \textbf{\begin{tabular}[c]{@{}c@{}}Phone Speaker\\ Location\end{tabular}} \\ \hline
        LG G3                                                            & \begin{tabular}[c]{@{}c@{}}Invensense\\ MPU-6500\end{tabular}     & 4-4000Hz                  & Back                                                                      \\ \hline
        \begin{tabular}[c]{@{}c@{}}Samasung Galaxy\\ Note 4\end{tabular} & \begin{tabular}[c]{@{}c@{}}Invensense\\ MPU-6515\end{tabular}     & 4-4000Hz                  & Back                                                                      \\ \hline
        \begin{tabular}[c]{@{}c@{}}Samsung Galaxy\\ S6\end{tabular}      & \begin{tabular}[c]{@{}c@{}}Invensense\\ MPU-6500\end{tabular}     & 4-4000Hz                  & Bottom Edge                                                               \\ \hline
    \end{tabular}}
\end{table}

 \begin{figure}[H]
     \centering
     \includegraphics[scale=0.3]{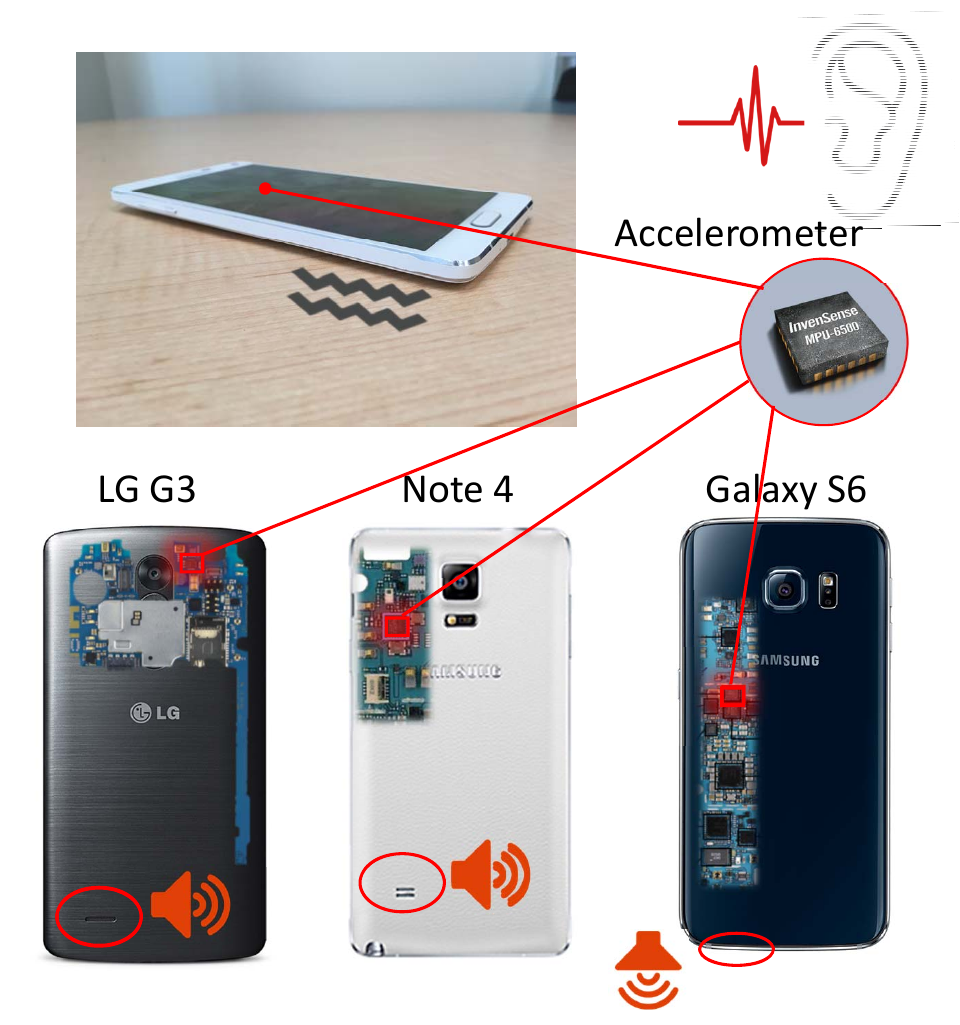}
     \vspace{-2mm}
     \caption{The speaker and the sensor positions on the smartphones of different vendors.}
     \label{fig:hardware_reason}
 \end{figure}
 
  \subsection{Accelerometer Response}
 \begin{figure}[H]
 	\begin{center}
 		\begin{tabular}{cc}
 			\includegraphics[scale=0.28]{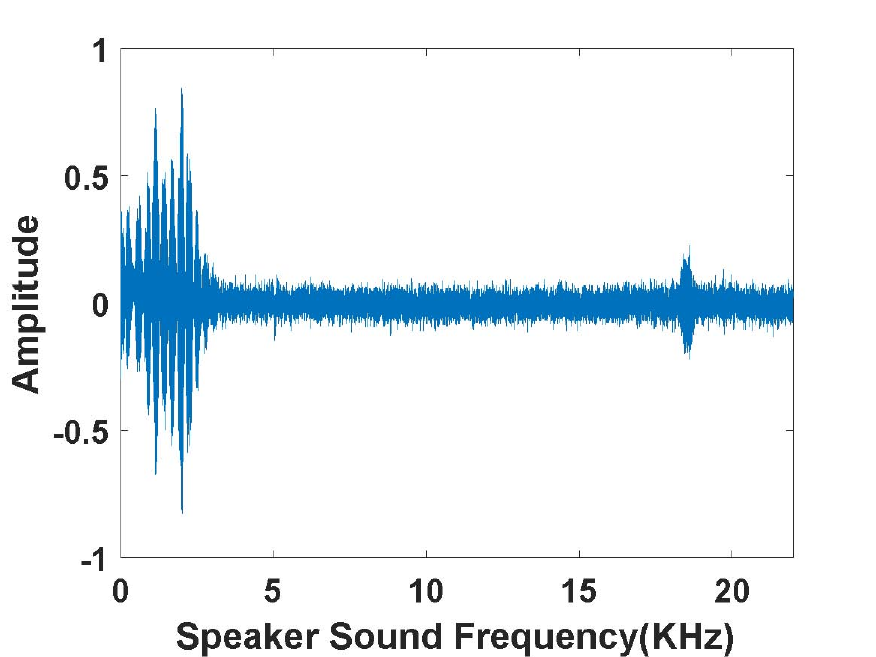}
 			&
 			\includegraphics[scale=0.28]{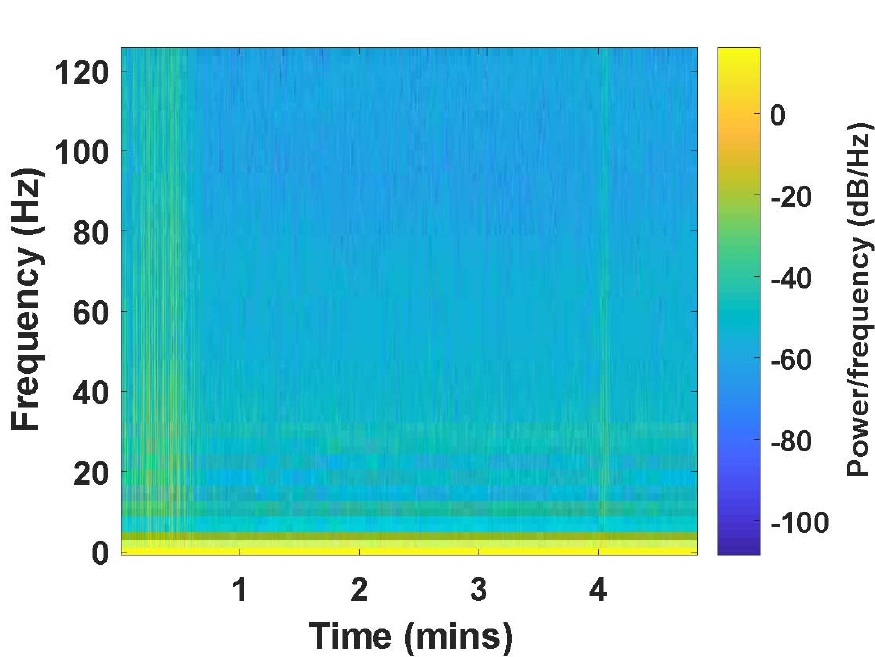}
 			\\
 			{\scriptsize(a) Amplitude for sound 0 - $22$kHz}&	
 			{\scriptsize(b) Spectrum for sound 0 - $22$kHz} 
 		\end{tabular}
 	\end{center}
 	\caption{Frequency response of the accelerometer along the z axis in response to a frequency-sweeping sound played by the smartphone's built-in loudspeaker.}
 	\label{fig:frequency_response1}
 \end{figure}

  \subsection{Accelerometer Response with Different Propagation Medium}
\begin{figure}[H]
	\begin{center}
		\begin{tabular}{cc}
			\includegraphics[scale=0.28]{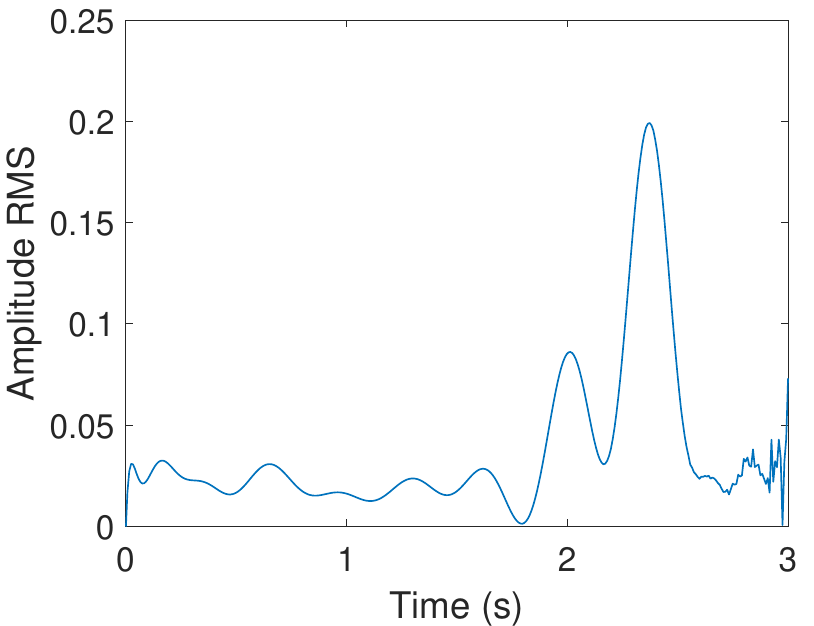}
			&
			\includegraphics[scale=0.28]{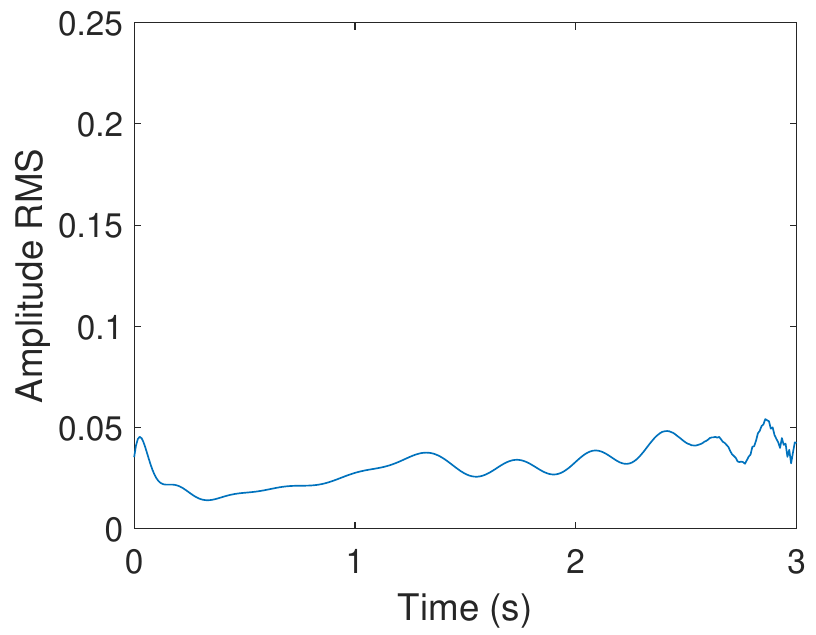}
			\\
			{\scriptsize(a) Smartphone body}&	
			{\scriptsize(b) Shared solid surface} 
		\end{tabular}
	\end{center}
	\caption{The RMS of the accelerometer's response to the two experimental settings: (1) Smartphone body: the phone's accelerometer captures the reverberations from the phone's own loudspeaker; and (2) Shared solid surface: the phone's accelerometer captures the vibrations from another phone's loudspeaker via the shared solid surface.}
	\label{fig:propagation_medium}
\end{figure}

 \subsection{Gyroscope Response}
 \label{sec:gyroRes}
 A gyroscope is a motion-sensing device used to measure device's angular velocity. 
 The main principle of the MEMS gyroscope is the Coriolis effect, which causes an object to exert a force when it is rotating. This force can be measured by a capacitive-sensing structure supporting the vibrating mass to determine the rate of rotation. Appendix Figure~\ref{fig:frequency_response_gyro} shows the gyroscope response to the $0-22$kHz frequency sweeping sounds from the built-in speaker. Gyroscope has observable response in the frequency range $8 - 9$kHz and $18 - 19$kHz and thus can capture some sound information in these frequency ranges. However, compared to accelerometer, gyroscope has a weaker response to the built-in loudspeaker's sound. In particular, the gyroscope shows subdued response in the frequency range $0-4$kHz (i.e., for audio sampled at $8$kHz), which is more often used in practical scenarios such as telephone calls and voice messages and the speech sound lies in this frequency range. 
 Given this property of gyroscope, we only focus on using the smartphone's accelerometer to capture the speech information.
 
 To verify this observation, we captured a single speaker's voice in both Gyrophone \cite{michalevsky2014gyrophone} setup and our proposed setup as described in Section \ref{sec:design} and implemented in Section \ref{sec:attack}. The gyroscope readings' spectrum from Gyrophone setup and the accelerometer readings' spectrum from \spyker\ setup are shown in Appendix Figure \ref{fig:specCompare}. We observed no indication of speech on Appendix Figure \ref{fig:specGyro} spectrum while we noticed the speech reverberations corresponding to word ``Oh'' around 3.5 second mark in Appendix Figure \ref{fig:specAccel} further validating our findings. We also further noted that Gyrophone setup involved a shared conducting medium that transferred the speech vibrations from the external loudspeaker to the smartphone's motion sensor. Thus, the capacity of motion sensors like gyroscope to sense these speech vibrations depends upon the nature of the shared surface. In contrast, \spyker\ setup detects speech reverberations, traveling within the smartphone's body, thus is independent of any such external causes.
 
 \begin{figure}[H]
\centering
	\begin{center}
		\begin{tabular}{cc}
			\includegraphics[scale=0.28]{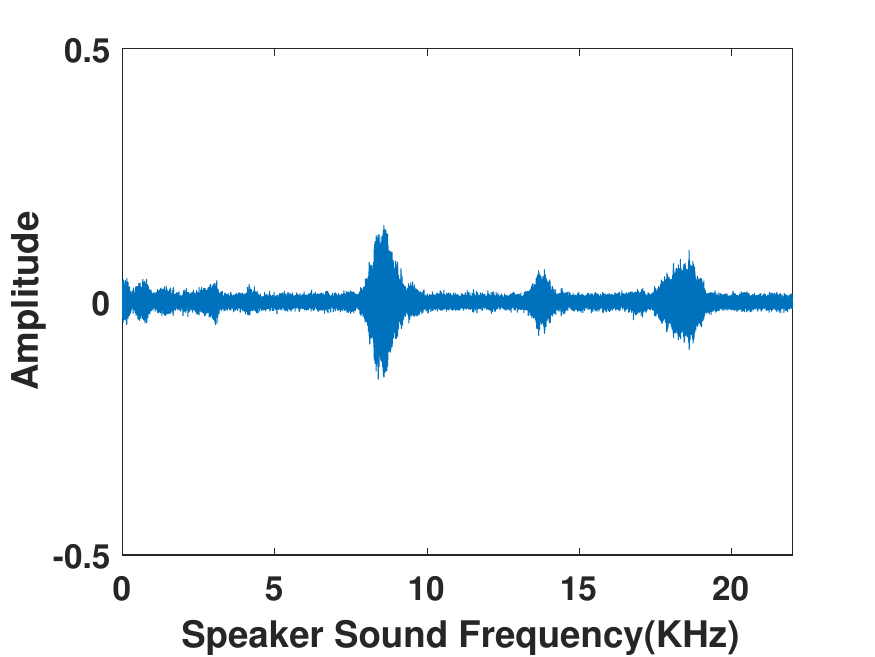}
			&
			\includegraphics[scale=0.28]{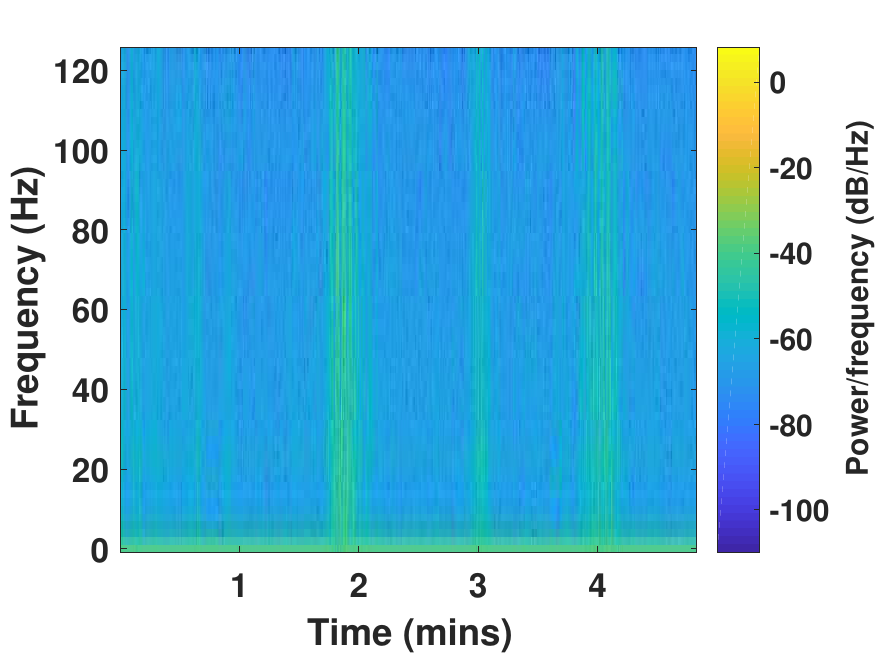}
			\\
			{\scriptsize(a) Gyroscope reading} &	
			{\scriptsize(b) Spectrogram} 
		\end{tabular}
	\end{center}
	\vspace{-2mm}
	\caption{Frequency response of the gyroscope to 0 - 22kHz frequency sweeping sound}
	\label{fig:frequency_response_gyro}
\end{figure}

 \begin{figure}[H]
\centering
\subfloat[Gyroscope readings' power density spectrum]{\includegraphics[width=0.45\columnwidth, height=3.5cm]{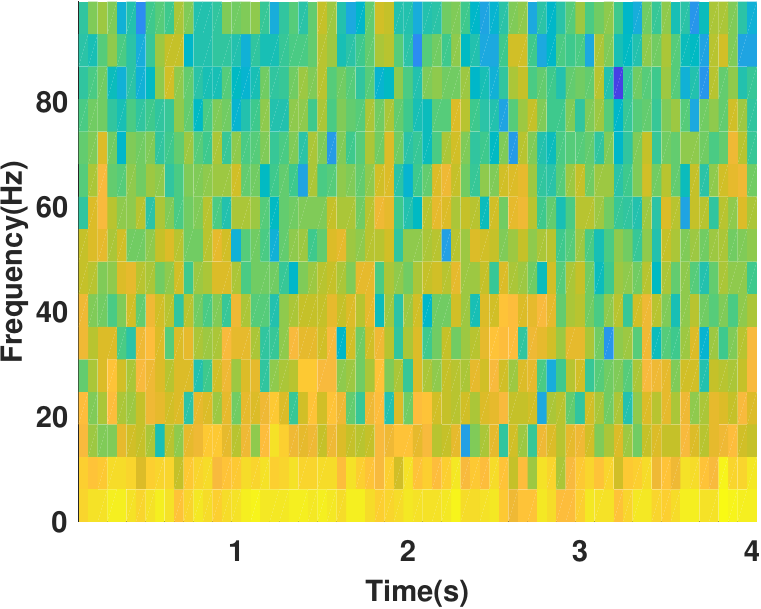}
\label{fig:specGyro}} \hspace{2mm}
\subfloat[Accelerometer readings' power density spectrum]{\includegraphics[width=0.45\columnwidth, height=3.5cm]{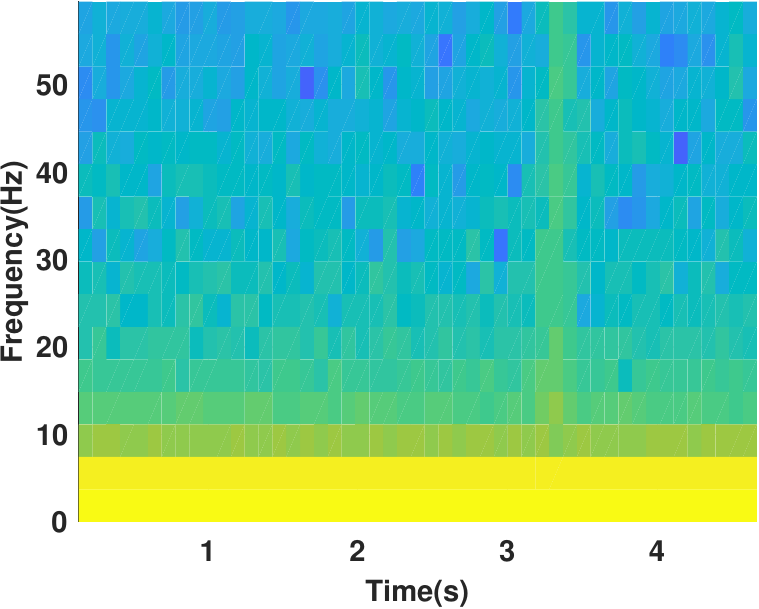}
\label{fig:specAccel}}
\vspace{-2mm}
\caption{Spectrum comparison (z axis) for the speaker ``MAE'' pronouncing the word ``Oh'' (TIDigits dataset) in \cite{michalevsky2014gyrophone} setup and \spyker\ setup.}
\label{fig:specCompare}
\end{figure}

\subsection{Evaluation of the Ear Piece Speaker}

\begin{figure}[H]
\centering
\includegraphics[scale=0.35]{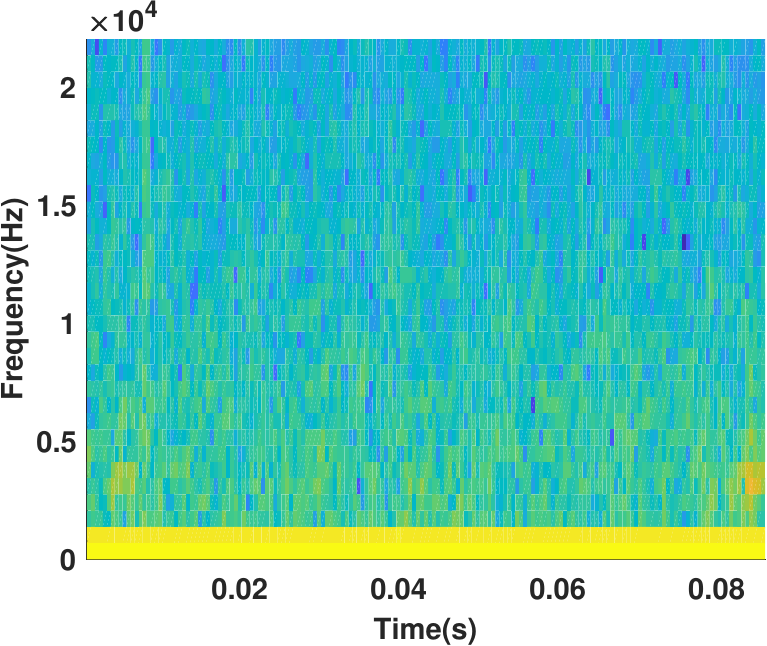}
\vspace{-2mm}
\caption{Spectrum of accelerometer (LG G3) with maximum call volume on the ear piece speaker. An incoming voice call was initiated where the caller uttered digits ``0'' to ``9'' and ``oh''.}
\label{fig:earLG}
\end{figure}


\pagebreak
\subsection{Comparison of Various Classifiers}
\vspace{-2mm}
\begin{figure}[H]
    \centering
    \subfloat[Gender classification (10-fold cross validation model)]{{\includegraphics[width=0.4\columnwidth,height=3cm]{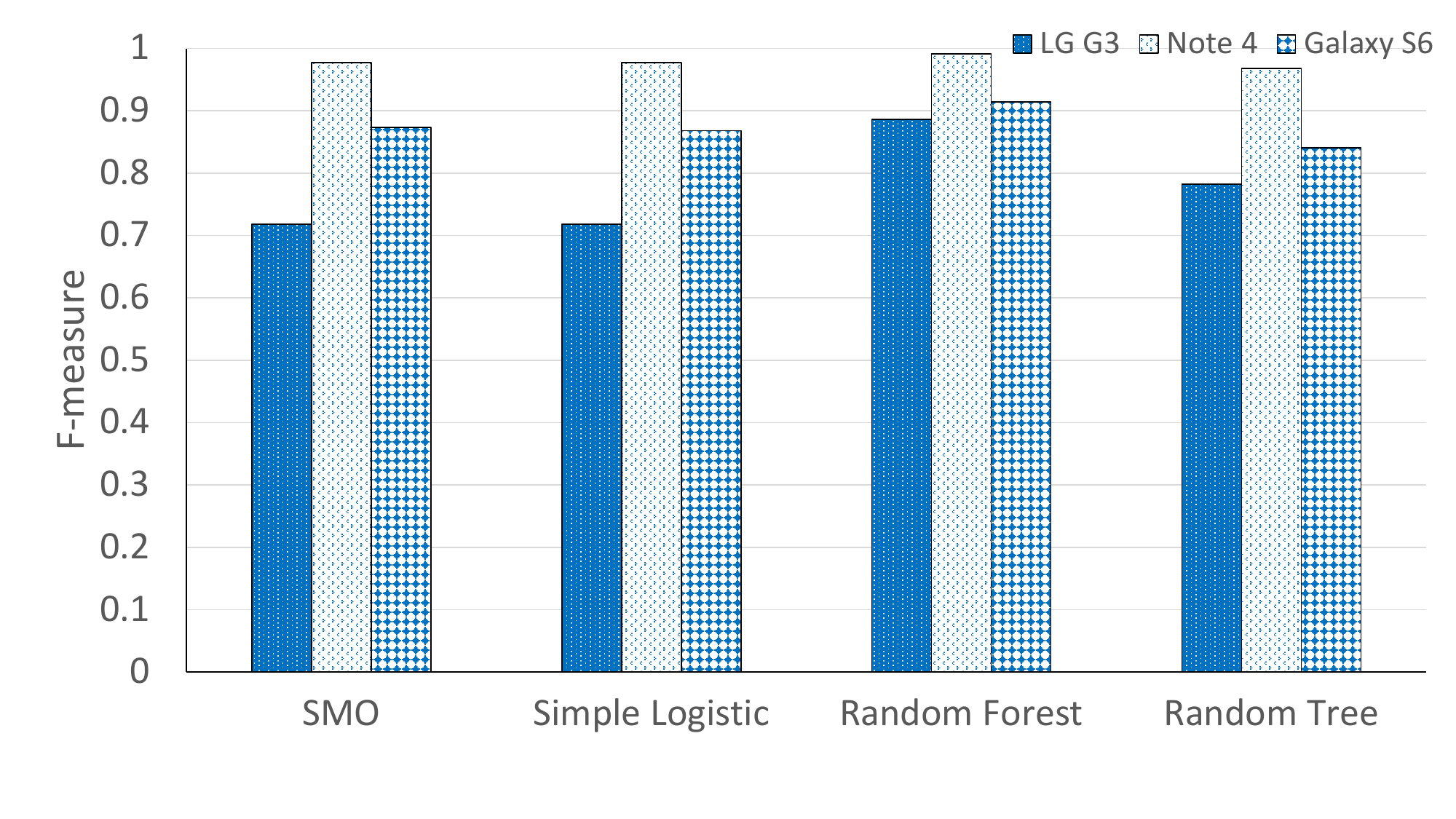}}
        \label{fig:HsTDGend10fold}}
    \subfloat[Gender classification (train-test model)]{{\includegraphics[width=0.4\columnwidth,height=3cm]{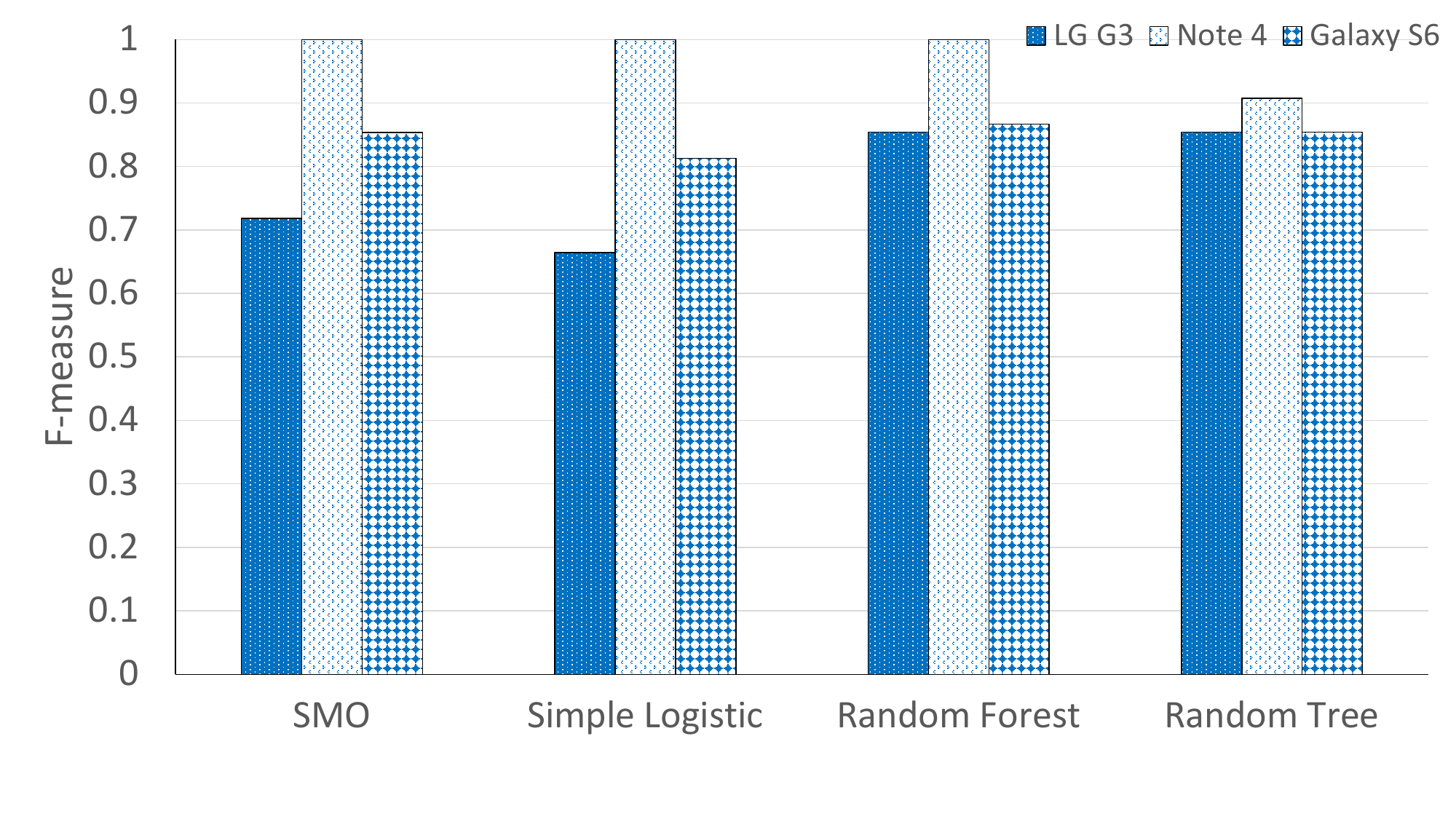}}
        \label{fig:HsTDGendtt}}\\
    \vspace{-2mm}
    \subfloat[Speaker classification (10-fold cross validation model)]{{\includegraphics[width=0.4\columnwidth,height=3cm]{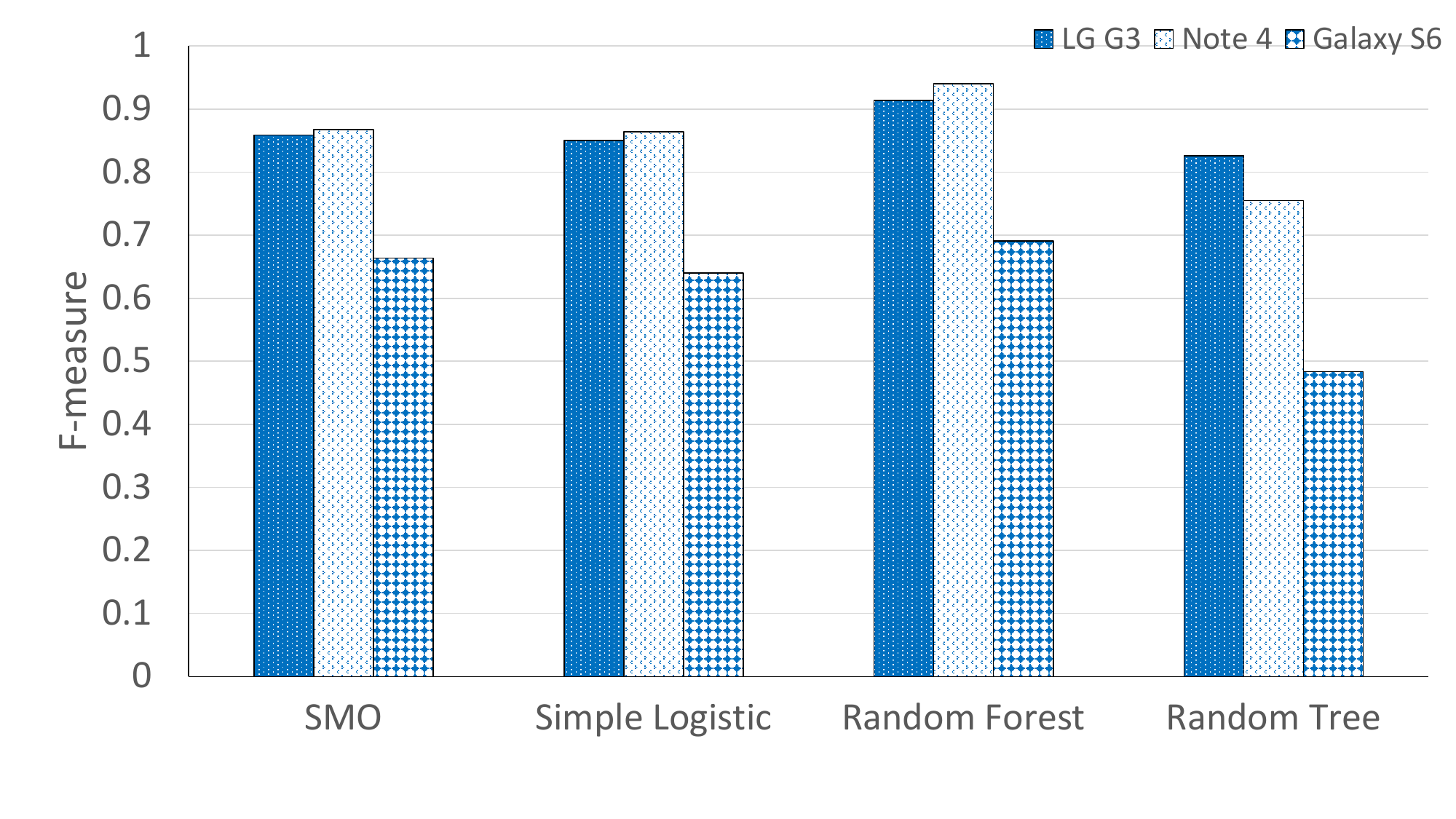}}
        \label{fig:HsTDSpk10fold}}
    \subfloat[Speaker classification (train-test model)]{{\includegraphics[width=0.4\columnwidth,height=3cm]{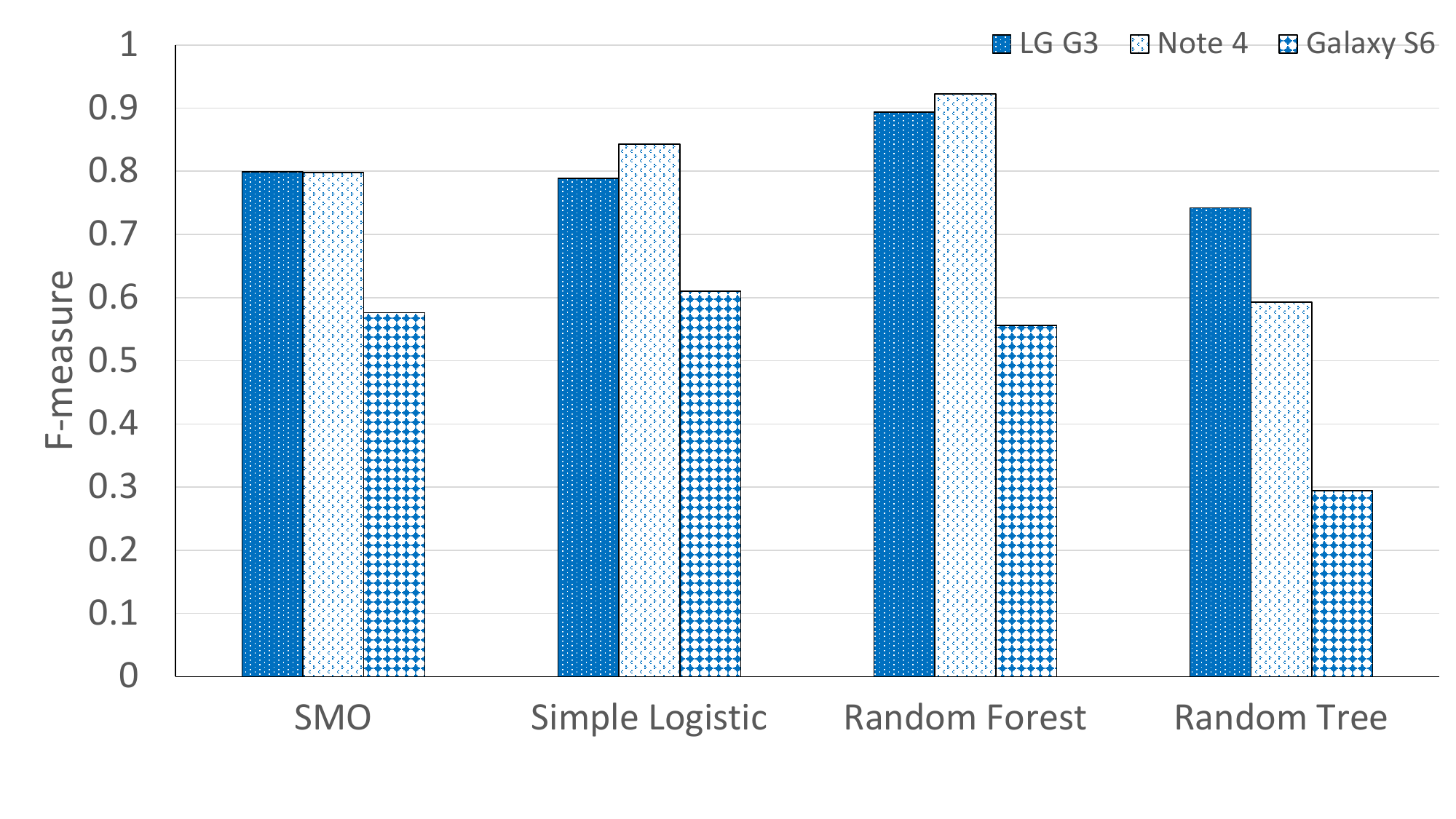}}
        \label{fig:HsTDSpktt}}
    \caption{Gender and speaker classification (10 speakers) for \phs\ setup using TIDigits dataset}
    \label{fig:PhsTD}
    \vspace{-3mm}
\end{figure}

\begin{figure}[H]
    \centering
    \subfloat[Gender classification (10-fold cross-validation model)]{{\includegraphics[width=0.4\columnwidth,height=3cm]{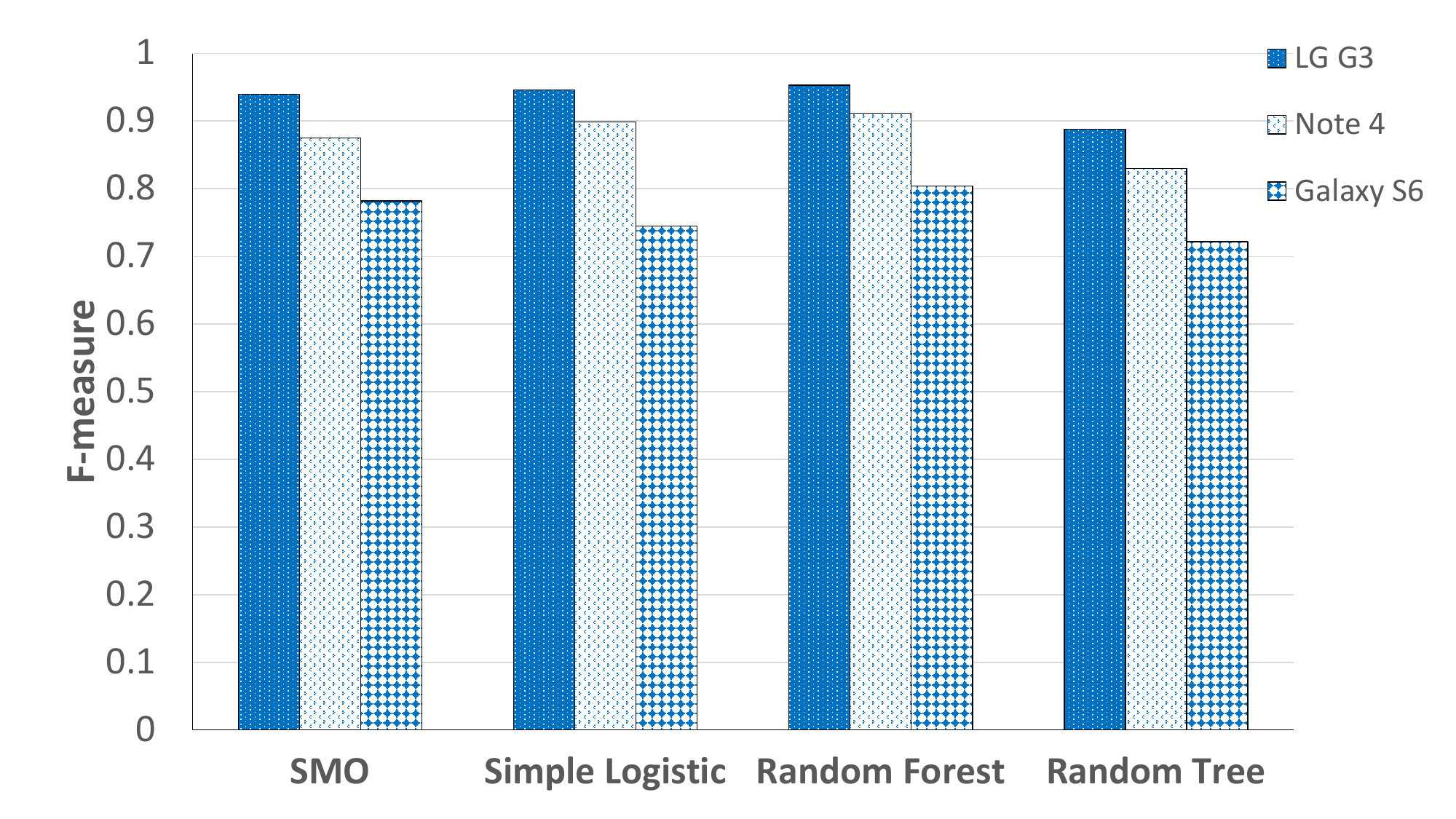}}
        \label{fig:HsAmGend10fold}}
    \subfloat[Gender classification (train-test model)]{{\includegraphics[width=0.4\columnwidth,height=3cm]{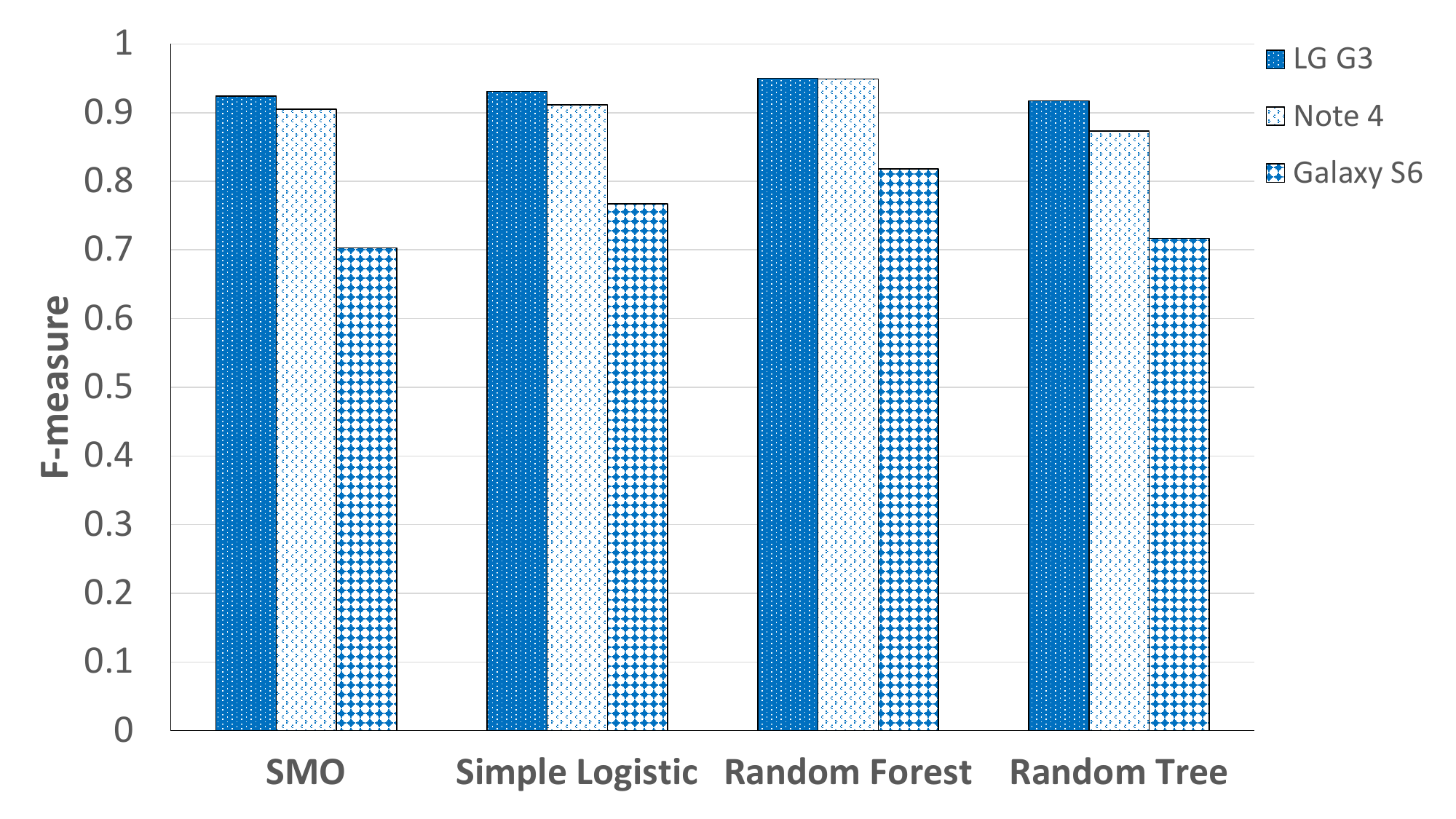}}
        \label{fig:HsAmGendtt}}\\
    \vspace{-2mm}
    \subfloat[Speaker classification (10-fold cross-validation model)]{{\includegraphics[width=0.4\columnwidth,height=3cm]{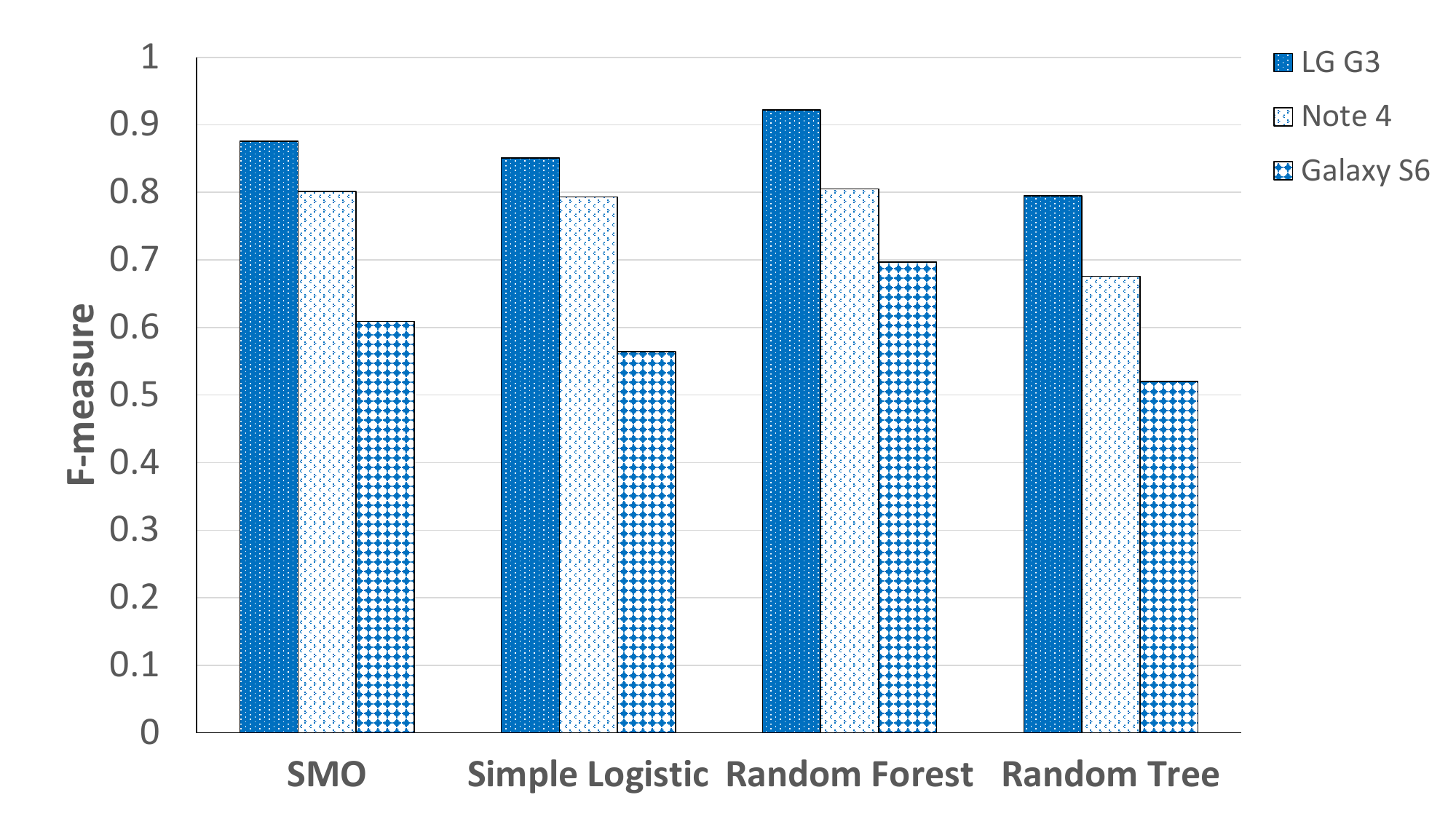}}
        \label{fig:HsAmSpk10fold}}
    \subfloat[Speaker classification (train-test model)]{{\includegraphics[width=0.4\columnwidth,height=3cm]{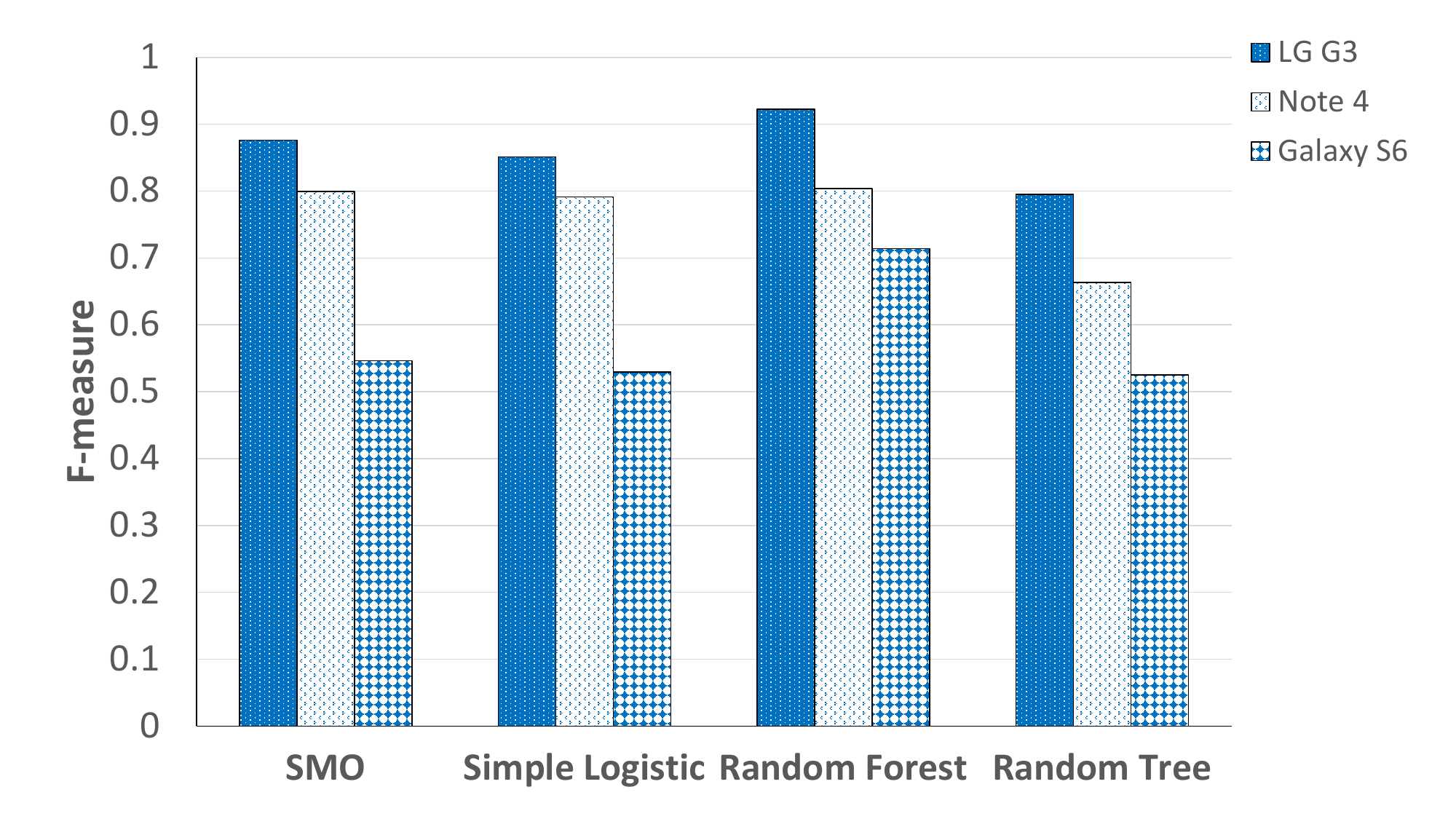}}
        \label{fig:HsAmSpktt}}
    \caption{Gender and speaker classification (10 speakers) for \phs\ setup using PGP words dataset}
    \label{fig:PhsAm}
\end{figure}

\subsection{Time-frequency Feature List}
\vspace{-2mm}
\begin{table}[H]
	\centering
	\caption{The time-frequency features calculated from accelerometer readings of X, Y and Z axis over a sliding window}
	\vspace{-2mm}
	\label{table:feature_list}
	\resizebox{\columnwidth}{!}{
		\begin{tabular}{|l|}
			\hline
			\textbf{Time Domain} \\ \hline
			Minimum; Maximum; Median; Variance; Standard deviation; Range \\ \hline
			CV: ratio of standard deviation and mean times 100 \\ \hline
			Skewness (3rd moment); Kurtosis (4th moment) \\ \hline
			Q1, Q2, Q3: first, second and third quartiles \\ \hline
			Inter Quartile Range: difference between the Q3 and Q1 \\ \hline
			Mean Crossing Rate: measures the number of times the signal crosses the mean value \\ \hline
			Absolute Area: the area under the absolute values of accelerometer signal \\ \hline
			Total Absolute Area: sum of Absolute Area of all three axis \\ \hline
			Total Strength: the signal magnitude of all accelerometer signal of three axis averaged of all three axis \\ \hline
			\textbf{Frequency Domain} \\ \hline
			Energy \\ \hline
			Power Spectral Entropy \\ \hline
			Frequency Ratio: ratio of highest magnitude FFT coefficient to sum of magnitude of all FFT coefficients \\ \hline
	\end{tabular}}
\end{table}

\subsection{Salient Features for Gender, Speaker, and Word Classification}
\vspace{-2mm}
 \begin{figure}[H]
	\centering
	\includegraphics[width=2.8in]{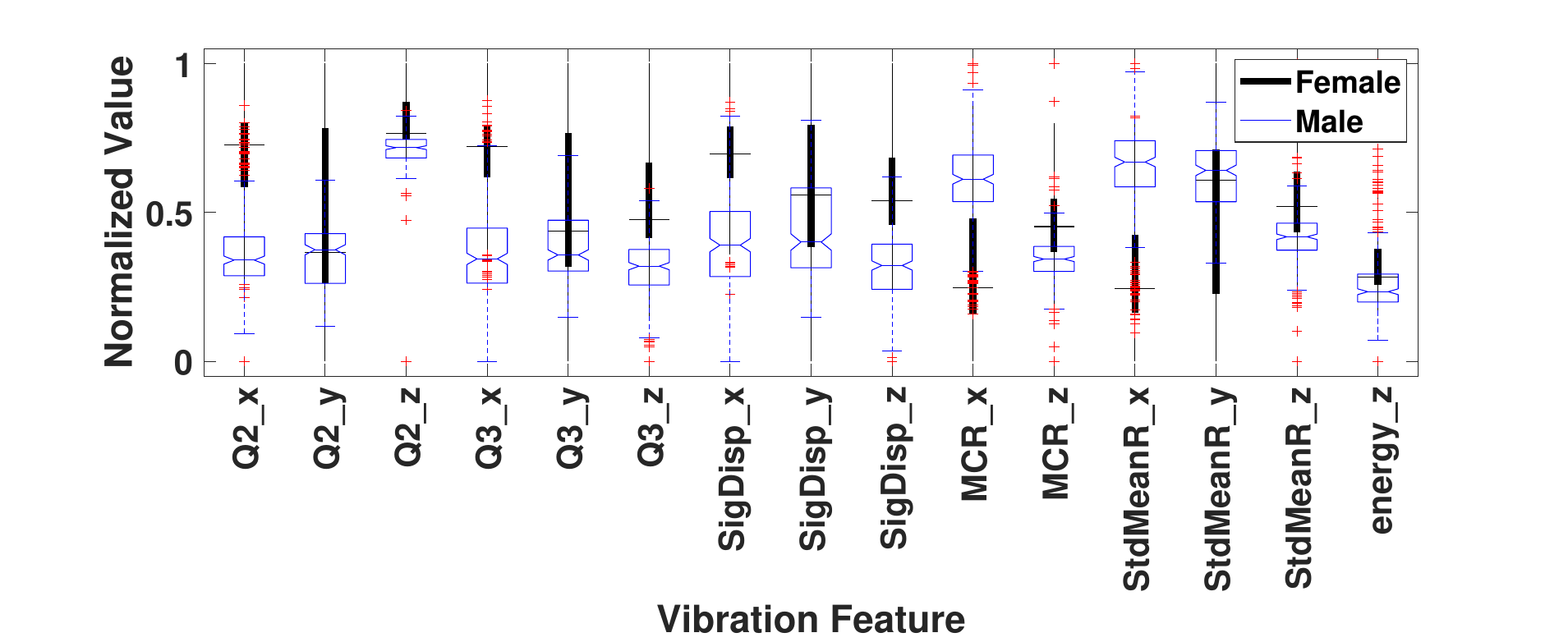}
	\vspace{-3mm}
	\caption{Salient time-frequency feature distributions for \gendCl.}
	\label{fig:spearphone_salient_features}
	\vspace{-5mm}
\end{figure}

\begin{figure}[H]
	\centering
	\includegraphics[width=\columnwidth]{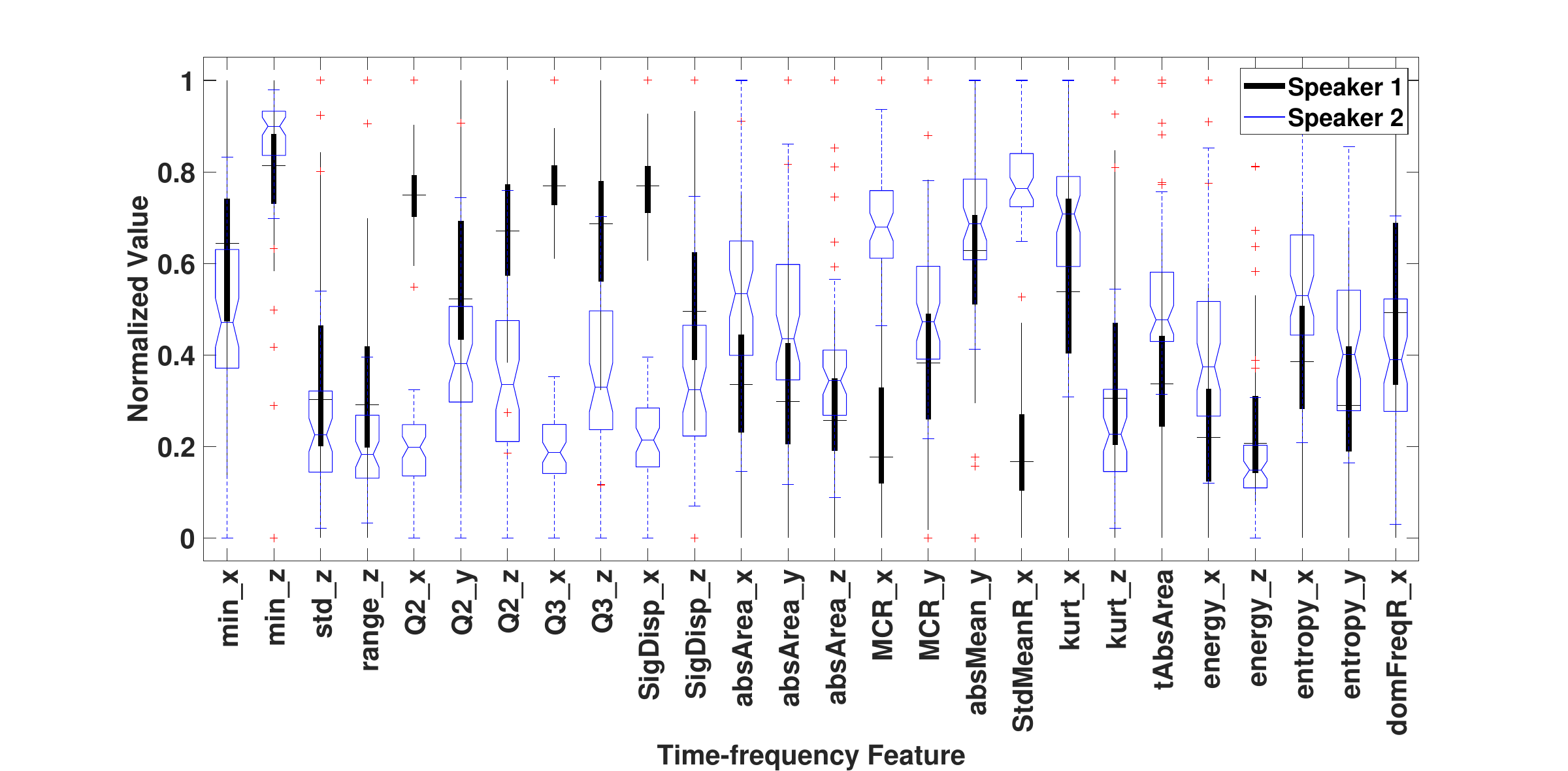}
	\caption{Illustration of the salient time-frequency features to differentiate speakers.}
	\label{fig:salient_feature_speaker}
\end{figure}

\begin{figure}[H]
	\centering
	\includegraphics[width=\columnwidth]{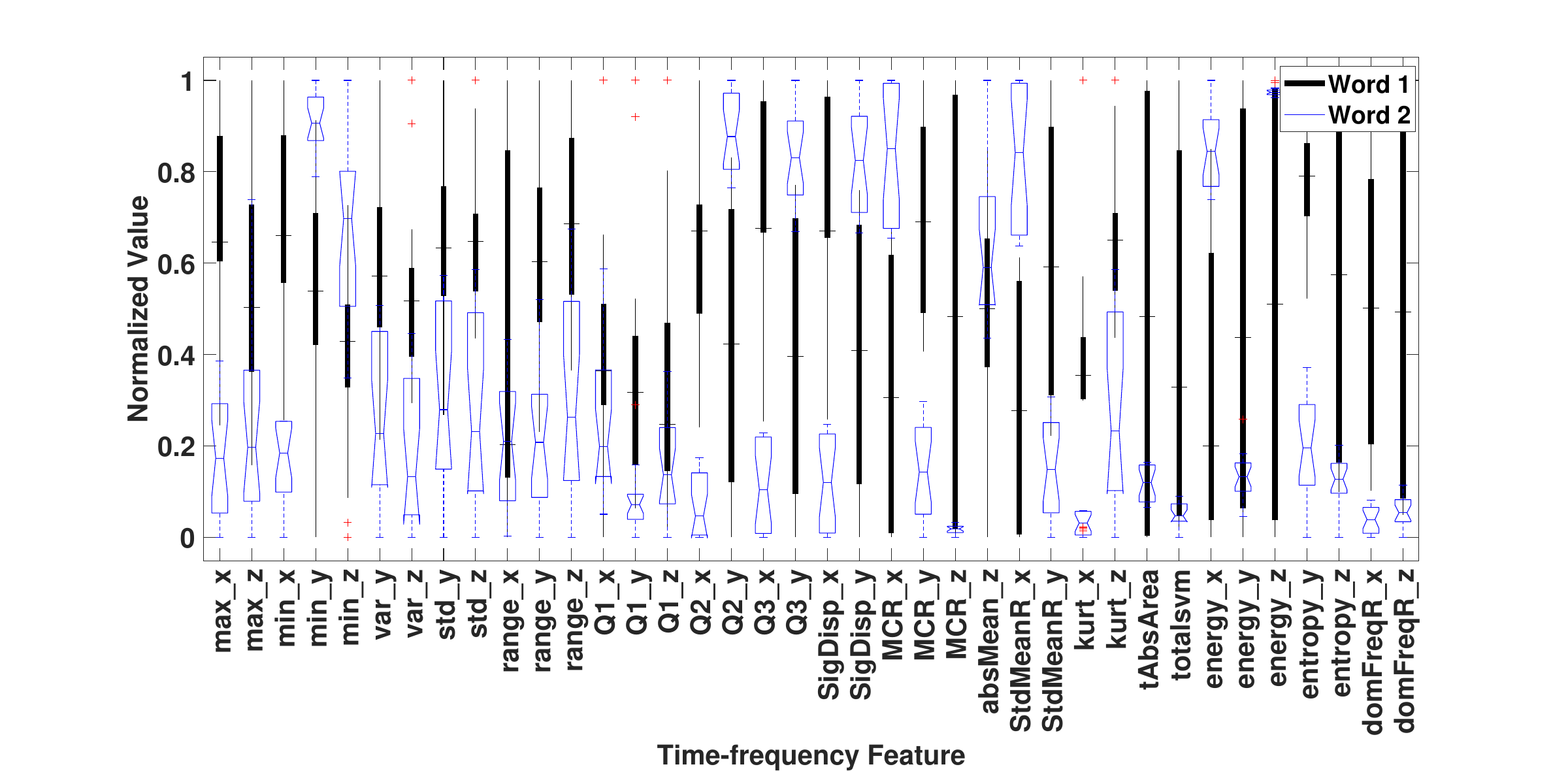}
	\caption{Illustration of the salient time-frequency features to differentiate words.}
	\label{fig:salient_feature_word}
\end{figure}

\end{document}